%% file: pap10.tex
\documentclass[12pt]{article}
\usepackage{epsf}
\usepackage{graphicx}
\usepackage{amsfonts}
\usepackage[mathscr]{eucal}

\def\href#1#2{#2}   
\newif\ifdraft
\draftfalse	  
\reversemarginpar   
\makeatletter
\let\mlabel=\label
\let\adkendequation=\endequation%
\def\endequation{\adkendequation\adklabel\global\@ignoretrue}
\let\adkendeqnarray=\endeqnarray%
\def\endeqnarray{\adkendeqnarray\adklabel\global\@ignoretrue}
\newbox\marglabbox
\def\adklabel{\ifvoid\marglabbox\else\marginpar{\unhbox\marglabbox}\fi}
\def\label#1{\ifdraft\ifmmode%
  \global\setbox\marglabbox=\hbox{\hfill\fbox{\tiny\verb*~#1~}}%
  \else\ifinner\else\marginpar{\hfill\fbox{\tiny\verb*~#1~}}%
  \fi\fi\fi \mlabel{#1}}
\makeatother
\ifdraft%
\fi
\def\bb{\mathbb}
\def\eusm{\mathscr}
\def\sqr#1#2{{\vcenter{\hrule height.#2pt
   \hbox{\vrule width.#2pt height#1pt \kern#1pt
      \vrule width.#2pt}
   \hrule height.#2pt}}}

\def\bsqr#1#2{{\vrule width #1pt height#2pt}}
\def\bsquare{{\mathchoice\bsqr66\bsqr66\bsqr33\bsqr33}}
\def\badbreak{\penalty1000}

\def\Trs{\mathop{\rm tr}}		    
\def\Trb{\mathop{\rm Tr}}		    
\def\identity{{\bb I}}			    
\def\rational#1#2{{\mathchoice{\textstyle{#1\over#2}}%
  {\scriptstyle{#1\over#2}}{\scriptscriptstyle{#1\over#2}}{#1/#2}}}
\def\half{\rational12}			    
\def\R{{\bb R}}				    
\def\Z{{\bb Z}}				    
\newcommand{\muhat}{{\hat \mu}}             
\newcommand{\chat}{{\hat c}}                
\newcommand{\Fhat}{{\hat F}}                
\newcommand{\cO}{{\cal O}}                  
\newcommand{\cP}{{\cal P}}                  
\newcommand{\cX}{{\cal X}}                  
\newcommand{\euS}{{\eusm S}}                
\newcommand{\euT}{{\eusm T}}                
\newcommand{\euV}{{\eusm V}}                
\newcommand{\hrho}{{\hat \rho}}             
\newcommand{\tdel}{{\tilde \delta}}         

\begin{document}

\begin{center}
{\Large{\bf Classical Limits of Scalar and Tensor Gauge}} \\
\vspace*{.1in}
{\Large{\bf Operators Based on the Overlap Dirac Matrix}} \\
\vspace*{.24in}
{\large{Andrei Alexandru$^1$, Ivan Horv\'ath$^2$ and Keh--Fei Liu$^2$}}\\
\vspace*{.24in}
$^1$The George Washington University, Washington, DC 20052\\
$^2$University of Kentucky, Lexington, KY 40506

\vspace*{0.15in}
{\large{Jul 4 2008}}

\end{center}

\vspace*{0.05in}

\begin{abstract}

  \noindent
  It was recently proposed by the second author to consider lattice 
  formulations of QCD in which complete actions, including the gauge
  part, are built {\em explicitly} from a given Dirac operator $D$. 
  In a simple example of such theory, the gauge action is proportional 
  to the trace of Ginsparg-Wilson operator $D$ chosen to define the quark 
  dynamics. This construction relies on the proposition that the classical 
  limit of lattice gauge operator $\Trs D(x,x)$ is proportional to 
  $\Trs F^2(x)$ (up to an additive constant). Here we show this for 
  the case of the overlap Dirac operator using both analytical and 
  numerical methods. We carry out the same analysis also for the tensor 
  component of $D$, which is similarly related to the field--strength 
  tensor $F$, and obtain results identical to our previous derivation 
  that used different approach. The corresponding proportionality constants 
  are computed  to high precision for wide range of the negative mass 
  parameter values, and it is verified that they are the same in finite 
  and infinite volumes.
\end{abstract}

\vspace*{0.10in}

\section{Introduction}

To define various versions of lattice QCD (LQCD), one usually follows a
standard route where the gauge part of the total continuum action
is treated independently from the part involving fermions. Indeed,
one normally writes down a candidate for lattice gauge action as an 
explicit function of link variables with appropriate symmetries,
such that its continuum limit on smooth backgrounds coincides with 
the continuum expression. On the other hand, the basic entity
in the construction of fermionic part is the lattice Dirac operator
$D$, itself a collection of functions in gauge variables (matrix elements).
In a standard treatment, one does not attempt to functionally relate gauge 
action to matrix elements of $D$. 

It has recently been suggested by one of us~\cite{Hor06B,Hor06C} that 
exploring lattice gauge theories with gauge and fermionic parts of 
the action explicitly related ({\em coherent LQCD}) could be beneficial 
for studies of QCD vacuum structure. This is particularly attractive
if $D$ is a chirally symmetric operator of Ginsparg--Wilson 
type.~\footnote{It is worth emphasizing that the notion of coherent LQCD 
can be well-defined even if $D$ is not chirally symmetric, assuming that 
a suitable function $f(D)$ is used to define required gauge operators. 
For example, the choice $f(D)=D^4$ produces coherent LQCD for arbitrary 
lattice Dirac operator $D$~\cite{Hor06B}.} One unusual feature of coherent 
formulations is the possibility of incorporating explicit relations between 
gauge and fermionic aspects of the theory. For example, it was suggested 
that QCD with $N_f$ quark flavors can be regularized via lattice formulation 
in which the usual gluon kinetic term is traded for additional quark 
flavor(s) whose mass controlls the gauge coupling 
(symmetric logarithmic LQCD)~\cite{Hor06B,Hor06C}. It remains to be seen
if one can gain theoretical or computational advantage by introducing 
interrelations of this type into the lattice--regularized theory.

The validity of various coherent versions of LQCD hinges on conjectures 
of {\em locality} and proper {\em classical limit} for scalar and pseudoscalar 
gauge densities associated with operator functions $f(D)$ used in 
the construction. If $D$ is a Ginsparg--Wilson operator, which we assume 
from now on, 
then these are expected to be technically non--trivial issues due to 
the fact that $D$ is necessarily a non--ultralocal operator in fermionic 
variables~\cite{nonultr}, and its dependence on gauge degrees of freedom is 
also expected to be generically non--ultralocal. 

In this paper, we discuss the issue of classical limit for coherent LQCD, where 
the gauge action is based on $\Trb D$. More precisely, we will focus on standard 
overlap operators $D^{(\rho,r)}$~\cite{Neu98BA} constructed from Wilson--Dirac 
matrix (negative mass $-\rho$, Wilson parameter $r$), and show that the 
classical limit of $\Trs D^{(\rho,r)}(x,x)$ is proportional to 
$\Trs F_{\mu\nu}(x)F_{\mu\nu}(x)$ up to an additive constant~\cite{Hor06B}. 
It should be emphasized that in formulations with $f(D)=D$ one does not have 
to deal with the issue of locality. Indeed, the required locality properties follow 
from locality of $D$, which is imposed to begin with.  For overlap Dirac operators 
the locality was studied e.g. in Refs.~\cite{ov_loc,Dra06A}, and the specific 
issue of locality in gauge variables was indirectly checked via reflection 
positivity considerations in Ref.~\cite{Hor05B}. 

In addition to justifying simple coherent LQCD as a valid regulator, the validity 
of conjectured classical limit for $\Trs D^{(\rho,r)}(x,x)$ will also make 
possible its use as a coherent scalar partner to overlap--based pseudoscalar 
density (topological density)~\cite{Has98A,NarNeu95,Kik98A,c_pscalar}. The associated 
noise reduction due to non--ultralocality is expected to be useful for obtaining 
refined information on the fundamental vacuum structure first discovered in 
overlap--based topological density~\cite{Hor02D,Hor03A}, as well as for 
conventional uses such as studies of glueball spectrum~\cite{Che05A}. 
Needless to say, for these purposes, it would 
in fact be desirable to have all interesting composite fields defined coherently. 
Important example is the field strength tensor $F_{\mu\nu}$, which can 
be constructed from tensor component of 
$D^{(\rho,r)}(x,x)$~\cite{Nie98A,Hor06B,KFL06A}.
Direct derivation has shown that in this case the required classical limit  
is indeed correct~\cite{KFL06A,KFL07A}. 

In principle, one could employ methods of Ref.~\cite{KFL07A} to derive 
the classical limit also in the scalar case considered here. However,  
the corresponding calculation is technically rather involved, and we thus take 
a different approach. In particular, the classical limit is first calculated 
for the class of constant gauge fields in infinite volume. This derivation is 
rigorous (even though technically it relies on a numerical evaluation of  
certain Brillouin zone integrals), and allows us to determine the 
corresponding proportionality constants $c^S(\rho,r)$ in a comparably simple 
calculation. Next, we show that our arguments also apply for constant fields in 
finite volume, at least for standard periodic and antiperiodic boundary 
conditions. Finally, we perform a direct numerical evaluation of the classical 
limit on generic non-constant backgrounds and find identical results again.
It is thus safe to conclude that the conjectured classical limit is indeed 
realized. Simultaneously with the scalar case, 
we apply our methods also to the tensor spinorial component and reproduce 
the results reported in Ref.~\cite{KFL07A}, thus cross-checking both approaches.
The proportionality constants $c^S$, $c^T$ and scalar subtraction constants
relevant for practical applications are given.

\section{Formulation of the Problem}
\label{sec:form}

Before we formulate the statements that we will be concerned with in this
paper, we wish to fix our notation and specify few conventions. 
\medskip

\noindent{\em (i) Continuum Gauge Fields.} The SU(3) gauge field 
$A_\mu(x)$ (traceless Hermitian 3 $\times$ 3 matrices) is related to 
the field--strength 
tensor via~\footnote{Note that this differs from conventions of
Ref.~\cite{Hor06B}, where anti--Hermitian gauge potentials were used
instead. The equations used here can be obtained from equations of
Ref.~\cite{Hor06B} via substitutions $A_\mu(x) \rightarrow i A_\mu(x)$,
$F_{\mu\nu}(x) \rightarrow i F_{\mu\nu}(x)$. The values of constants
$c^S$, $c^T$ are the same in both conventions.}
\begin{equation}
      F_{\mu \nu}(x) \,\equiv\, 
      \partial_\mu A_\nu(x) - \partial_\nu A_\mu(x) + 
      i\,[\, A_\mu(x), A_\nu(x) \,] 
   \label{eq:6}  
\end{equation} 
while the covariant derivative acts as
\begin{equation}
   D_\mu \,\phi(x) \,=\, (\partial_\mu + i A_\mu(x)) \,\phi(x)
   \qquad\quad
   [D_\mu,D_\nu] \, \phi(x) \,=\, i F_{\mu \nu}(x) \, \phi(x)
   \label{eq:11}
\end{equation}

\noindent {\em (ii) Classical Fields.} By classical continuum fields on $\R^4$
(or on a torus) we mean gauge potentials $A_\mu(x)$ smooth (differentiable 
arbitrarily many times) almost everywhere. If classical backgrounds contain
singularities, the classical continuum limits are assumed to be taken at its 
non--singular points. While our conclusion is expected to be valid for all 
classical fields (due to the locality of operators involved), in what follows 
we will only consider classical fields that are smooth everywhere. This avoids 
various inessential technical complications related to transcribing singular
fields on to the lattice.
\smallskip

\noindent {\em (iii) Transcription of Classical Fields to Hypercubic Lattice.}
Hypercubic lattice is superimposed on $\R^4$ via the correspondence 
$x\equiv na$ ($x\in \R^4$, $n\in \Z^4$), where $a$ is the lattice spacing. Smooth
continuum field is then transcribed to the lattice field via 
\begin{equation}
   U_{n,\mu}(a) \;\equiv\; 
   \cP \exp\Bigl(i a \int_0^1 ds\, A_\mu(an+(1-s)a\muhat)\Bigr)
   \label{eq:13}
\end{equation}
where $\cP$ is the path ordering symbol and $\muhat$ is a unit vector in 
direction $\mu$. 
\smallskip

\noindent {\em (iv) Overlap Operators.}
Standard overlap Dirac operators are defined by~\cite{Neu98BA} 
\begin{equation}
   D^{(\rho,r)}=\rho \, [\,1 \,+\, \cX (\cX^\dagger \cX)^{-\half}\,]  \qquad\quad
   \cX=D_W-\rho   \qquad\quad  \rho \in (0,2r) \qquad\quad r>0
   \label{eq:14.1}
\end{equation}
where $D_W\equiv 4r \, \identity \; - \half K$ is the massless Wilson--Dirac 
operator and $K$, the Wilson's hopping matrix with Wilson parameter $r$, is given by
\begin{equation}
   K_{n,m}\,=\, \sum_{\mu} \,(r-\gamma_{\mu})\, U_{n,\mu}\, \delta_{n+\mu,m} \,+\,
                         (r+\gamma_{\mu})\, U^\dagger_{n-\mu,\mu} \,\delta_{n-\mu,m} 
   \label{eq:14.3}
\end{equation}
For calculations in this paper, it is convenient to work with the rescaled
matrix $\cX$ namely
\begin{equation}
    \cX \;\longrightarrow\; X \,\equiv\, 2 \kappa \cX  \,=\, 
    \identity - \kappa K
    \qquad\qquad
    \kappa \,\equiv\, \frac{1}{8r-2\rho}
   \label{eq:14.4}
\end{equation}
The form of the overlap operator (\ref{eq:14.1}) in terms of $\cX$ is 
identical to that in terms of $X$.
\smallskip

\noindent {\em (v) Classical Limit Conjecture.} Our main goal in this paper is to 
support the Conjecture C3 of Ref.~\cite{Hor06B} for the specific case of standard
overlap Dirac operators. In the infinite volume, one can state this as 
follows.~\footnote{The subscripts ``s'', ``c'' in expressions that follow 
refer to ``spin'' and ``color'' respectively. Thus, for example, 
$\identity_c$ denotes the identity matrix in color space ($3\times 3$ matrix)
and ${\Trs}_{sc}$ denotes the trace in both spin and color.} 

\medskip

\noindent {\bf Conjecture 1.} {\em Let $A_\mu(x)$ be arbitrary smooth  {\rm SU(3)} 
gauge potentials on $\R^4$. If $U(a) \equiv \{\,U_{n,\mu}(a)\,\}$ is the transcription 
of this field to the hypercubic lattice with classical lattice spacing $a$, and 
$\identity \equiv \{\,U_{n,\mu} \rightarrow \identity_c\,\}$ is the free gauge 
configuration then 
\begin{equation}
   {\Trs}_{sc}\, \Bigl(\, D_{0,0} ( U(a) ) \,-\, 
                          D_{0,0} ( \identity )\, \Bigr) 
   \;=\;
   c^S  \,a^4 \, {\Trs}_c \, F_{\mu\nu}(0) F_{\mu\nu}(0) \;+\; \cO(a^6)
   \label{eq:14.5}
\end{equation} 
where $D\equiv D^{(\rho,r)}$ is the overlap Dirac operator and $F_{\mu\nu}(x)$ is 
the field--strength tensor associated with $A_\mu(x)$.
The constant $c^S\equiv c^S(\rho,r)$ is non--zero and independent of $A_\mu(x)$ at 
fixed $\rho$ and $r$.}

\medskip

\noindent We will also discuss classical limits for tensor components 
of $D^{(\rho,r)}$. The corresponding statement is obtained from the above by 
replacing Eq.~(\ref{eq:14.5}) with~\cite{Hor06B, KFL07A}
\begin{equation}
   {\Trs}_s\, \sigma_{\mu\nu} \, D_{0,0}(U(a)) 
   \;=\; c^T\, a^2\, F_{\mu\nu}(0) \;+\; \cO(a^4)
   \label{eq:14.6}
\end{equation} 
where $\sigma_{\mu\nu}\equiv \frac{1}{2i}[\gamma_\mu,\gamma_\nu]$.~\footnote{
We use Hermitian $\gamma$--matrices $\gamma_\mu^\dagger = \gamma_\mu$ 
throughout this paper.}  
\medskip

\noindent We emphasize that the above statements (with identical proportionality
constants $c^S(\rho,r)$, $c^T(\rho,r)$) are also expected to be true in a finite volume. 
We will verify this here for the practically relevant case of a torus. In what 
follows, we will frequently skip explicitly denoting dependences of operators/constants 
on parameters $\rho$ and $r$, but they are implicitly understood.

\section{Constant Fields in Infinite Volume}
\label{sec:const_fields}

In this section we establish that the leading term of Eq.~(\ref{eq:14.5}) is indeed
correct for the class of constant gauge fields $A_\mu(x)\equiv A_\mu$. This is
the simplest class of classical fields leading to non-zero field-strength tensor
(for non-Abelian gauge fields), while the resulting simplifications make the calculation 
reasonably manageable. 

For constant classical backgrounds the transcribed lattice field at lattice spacing
$a$ has the simple form~\footnote{It is worth mentioning at this point that one can 
use perturbation theory techniques to obtain expansions 
needed~\cite{Kik98A,Con07A,Ale00A,Ada07}. Here we proceed without reference
to weak coupling perturbation theory, as done in~\cite{c_pscalar,KFL07A}.}
\begin{equation}
   U_{n,\mu}(a) \;=\; \exp ( i a A_\mu )
   \label{eq:3.2}
\end{equation}
and the overlap Dirac matrix is translation invariant $D_{n,m}=D_{n-m,0}$.
The calculation of classical limit simplifies in the Fourier space, where 
the operator is diagonal in space--time indices. For arbitrary translation
invariant operator $O_{n,m}$ we define the diagonal Fourier image 
$O(k)$ via
\begin{equation}
    O(k) \,\equiv\, \sum_n e^{-i(n-m)k} \,O_{n,m}   \qquad\quad 
    O_{n,m} \,=\, \frac{1}{(2\pi)^4} \int d^4 k \,e^{i(n-m)k} \,O(k)
    \label{eq:3.4}
\end{equation}
where the integration over momentum variables runs through the Brillouin zone.
The definition of $O(k)$ requires the convergence of the above infinite sum, 
for which it is sufficient that $O$ is local. For arbitrary constant gauge field 
$A_\mu$ the overlap Dirac operator $D$ in transcribed lattice background $U_\mu(a)$ 
is guaranteed to be local~\cite{ov_loc} at sufficiently small classical lattice 
spacing $a$, and thus its Fourier transform is well defined.

To evaluate $D(k)$, we start from the expression for $D$ in terms of matrix $X$ 
(see Eq.~(\ref{eq:14.4}))
\begin{equation}
    \frac{1}{\rho} D \,=\, \identity + X \frac{1}{\sqrt{X^\dagger X}}
    \label{eq:3.6}
\end{equation}
The Fourier transform $X(k)$ of $X$ can be found straightforwardly and is given by
\begin{equation}
   X(k) \,=\, \identity_{sc} \,-\, 
              2\kappa r \sum_\mu \cos \Bigl(a A_\mu + k_\mu \Bigr)  \,+\, 
          i\, 2\kappa \sum_\mu \gamma_\mu \sin \Bigl( a A_\mu+k_\mu \Bigr)
   \label{eq:3.8}
\end{equation}
Moreover, the Hermitian matrix $B \equiv X^\dagger X$ is strictly positive--definite
for sufficiently small $a$ in arbitrary constant 
background~\cite{ov_loc}.~\footnote{The gap in the spectrum of $B$ tends to zero 
when excluded boundary values $\rho=0$ and $\rho=2r$ are approached.} Consequently,
the Fourier image of $B^{-1/2}$ is simply given by $(X^\dagger(k) X(k))^{-1/2}$,
and thus
\begin{equation}
    \frac{1}{\rho} D(k) \,=\, \identity_{sc} + X(k) \frac{1}{\sqrt{X^\dagger(k) X(k)}}
    \label{eq:3.10}
\end{equation}

\subsection{Expansion in Lattice Spacing}
\label{sec:expansion}

For the purposes of Conjecture 1, we are interested in the Taylor expansion (in $a$) of 
\begin{equation}
    \frac{1}{\rho} D_{0,0} \;=\; 
    \frac{1}{\rho}\, \frac{1}{(2\pi)^4} \int d^4 k \,D(k) \;=\;
    \identity_{sc} + 
    \frac{1}{(2\pi)^4} \int d^4 k \,X(k) \frac{1}{\sqrt{X^\dagger(k) X(k)}}
    \label{eq:3.12}
\end{equation}
which can be obtained term by term from the Taylor expansion of $X(k)B^{-1/2}(k)$ due 
to its analyticity in both $k$ and $a$ for sufficiently small $a$.

To fix the notation that we will use for various operators, we generically write
\begin{equation}
     O(k,a) \;=\; \sum_{n=0}^\infty O_n(k)\, a^n  \;\equiv\; O_0(k) + \delta O(k,a)
   \label{eq:3.14}
\end{equation}
where the dependence of $O(k)$ on $a$ will usually not be shown explicitly. The leading
terms for $X(k)$ and $B(k)$ are given by~\footnote{Note that in the formulas that follow
we denote $\identity_{sc}$ simply as ``1'' with spin--color structure understood.} 
\begin{equation}
   X_0(k) \,=\, 1 \,-\, 
              2\kappa r \sum_\mu \cos k_\mu  \,+\, 
          i\, 2\kappa \sum_\mu \gamma_\mu \sin k_\mu
   \label{eq:3.16}
\end{equation}
and 
\begin{equation}
   B_0(k) \,=\,  \Bigl( 1 - 2\kappa r \sum_\mu \cos k_\mu \Bigr)^2
                 \,+\, 4\kappa^2 \sum_\mu \sin^2 k_\mu
   \label{eq:3.18}
\end{equation}
There are two points to note here. {\em (i)} $B_0(k)$ is proportional to identity and 
is positive-definite for all $k$, except when $\kappa r =1/8, 1/4$ which corresponds 
to excluded boundary values of the $\rho$--parameter $0$, $2r$. This means that $B_0(k)$ 
can be commuted and inverted freely in the expressions that follow. 
{\em (ii)} The matrices $\delta X(k)$ 
and $\delta B(k)$ have norms that can be made arbitrarily small by lowering 
the lattice spacing sufficiently, and can thus serve as ``perturbations'' in matrix 
expansions around $X_0(k)$ and $B_0(k)$.   

The expansion of $X(k) B^{-1/2}(k)$ in lattice spacing, needed in (\ref{eq:3.12}), can 
now proceed via expanding $X(k)$ and $B^{-1/2}(k)$ separately and then combining the 
results. In case of $X(k)$, the expansion is obtained directly from 
(\ref{eq:3.8}), while for $B^{-1/2}(k)$ we use
\begin{equation}
    B^{-\half} \,=\, (B_0+\delta B)^{-\half} \,=\,
    B_0^{-\half} \, (1+ \tdel B)^{-\half}    \qquad\qquad
    \tdel B \,\equiv\, B_0^{-1}\delta B      
    \label{eq:3.20}
\end{equation}
Properties {\em (i)} and {\em (ii)} guarantee that $\tdel B$ is a valid perturbation
for the expansion of the inverse square root, yielding
\begin{equation}
   B^{-\half} \;=\;  B_0^{-\half}\, \Bigl[\,1 \,-\, 
   \frac{1}{2} \, \tdel B \,+\,
   \frac{3}{8} \, (\tdel B)^2 \,-\,
   \frac{5}{16} \, (\tdel B)^3 \,+\,
   \frac{35}{128} \, (\tdel B)^4  \,\Bigr]
   \;+\; \cO \Bigl((\tdel B)^5\Bigr) \,
   \label{eq:3.22}
\end{equation}
Note that we have to keep all the terms up to the 4-th order since 
we are eventually interested in $\cO(a^4)$ contribution and 
$\tdel B$ is $\cO(a)$. Indeed, the explicit formula for $B(k)$ in terms of its
Clifford decomposition is given by
\begin{equation}
     B(k) \,=\, \identity_s \times \euS(k)            \,+\,
                \gamma_\mu \times \euV_\mu(k)         \,+\,
                \sigma_{\mu\nu} \times \euT_{\mu\nu}(k) 
     \label{eq:3.24}
\end{equation}
where 
\begin{equation}
   \euS(k) \;=\; \Bigl(\, 1-2\kappa r\sum_\mu \cos(a A_\mu+k_\mu) \,\Bigr)^2 \,+\,
                     4\kappa^2\sum_\mu \sin(a A_\mu+k_\mu)^2
   \label{eq:3.26} 
\end{equation}
\begin{equation}
   \euV_\mu(k) \;=\; i 4 \kappa^2 \sum_\nu\, 
   \Bigl[\,\sin(a A_\mu+k_\mu), \cos(a A_\nu+k_\nu)\,\Bigr] \qquad\quad
   \label{eq:3.28}
\end{equation}
\begin{equation}
    \euT_{\mu\nu}(k) \;=\; i 2 \kappa^2 \,
    \Bigl[\,\sin(a A_\mu+k_\mu), \sin(a A_\nu+k_\nu)\,\Bigr]
    \label{eq:3.30}
\end{equation}
and $[,]$ denotes the commutator. From this one can inspect directly that 
while $\euV_\mu$ and $\euT_{\mu\nu}$ are $\cO(a^2)$, the scalar part
contains both the constant and the linear term, and thus $\tdel B$ is 
indeed $\cO(a)$.

The equations of this subsection together with Eq.~(\ref{eq:3.8}) define 
the procedure of evaluating various classical limits completely. The calculation
is straightforward, but turns out to be technically still quite involved if one
wants to obtain a complete expansion of $D(k)$ (and thus of $D_{0,0}$) all the 
way up to order $a^4$.

\subsection{Sample Computation -- First Order}
\label{sec:first_order}

To see the characteristic nature of the computations involved, let us 
now expand $D(k)$ up to order $a$. For this purpose, we will need 
expansions of $X(k)$ and $B(k)$ up to $\cO(a)$. We find
\begin{equation}
   X(k) \;=\; X_0(k) \;+\; a\, \sum_\mu A_\mu \,\underbrace{
    2\kappa \, (\,r \sin k_\mu  \,+\, i \gamma_\mu \cos k_\mu \,)}_{g_\mu(k)} 
   \;+\; \cO(a^2)
    \label{eq:3.32}
\end{equation} 
and 
\begin{equation}
   B(k) \;=\; B_0(k) \,+ \, a \sum_\mu A_\mu
   \underbrace{ \Bigl[\, 4\kappa r(1-2\kappa r\sum_\nu \cos k_\nu) \sin k_\mu
   \,+\, 8\kappa^2\sin k_\mu \cos k_\mu \,\Bigr]}_{f_\mu(k)} \,+\, \cO(a^2)
    \label{eq:3.34}
\end{equation}
Note that $f_\mu(k)$ comes entirely from $\euS(k)$.
Using the expansion (\ref{eq:3.22}) up to $\cO(\tdel B)$ we then obtain
$D(k) = D_0(k) + a D_1(k) + \cO(a^2)$ where
\begin{equation}
   \frac{1}{\rho} \,D_0(k) \;=\; 1 \,+\, 
   B_0(k)^{-1/2} \, X_0(k) 
   \label{eq:3.36}
\end{equation}
and
\begin{equation}
   \frac{1}{\rho}\, D_1(k) \;=\; \sum_\mu A_\mu
   \underbrace{ B_0(k)^{-3/2}\, 
   \Bigl[\, g_\mu(k)B_0(k) \,-\, \half \, f_\mu(k) X_0(k) \,\Bigr]
   }_{h_\mu(k)}
   \label{eq:3.38}
\end{equation}
The leading term $D_0(k)$ is the Fourier transform of the free overlap
operator as expected. 

Let us now focus on the first--order term which will determine (after 
inverse Fourier transform) the first--order term of $D_{0,0}$. Using
transformation properties, one can argue that this first--order term
should in fact vanish. Indeed, $D_{0,0}$ is a gauge covariant operator 
and this covariance has to be preserved order by order in the Taylor 
expansion. Thus, the coefficient of the linear term is expected to be 
a dimension one (continuum) gauge covariant operator (spin-color 
function of the gauge field), but such operator does not exist. 
On the other hand, the result (\ref{eq:3.38}) shows that $D_1(k)$ 
is not identically zero. This does not imply any contradiction as long 
as the associated inverse Fourier expression vanishes. One can see that 
this is indeed the case by evaluating $h_\mu(k)$ of Eq.~(\ref{eq:3.38}) 
explicitly, yielding
\begin{equation}
   h_\mu(k) \;=\; h_\mu^A (k) \,+\, 
                  i \gamma_\mu \, h_\mu^S (k) 
   \label{eq:3.40}
\end{equation} 
where $h_\mu^A (k)$ is antisymmetric with respect to $k_\mu \to -k_\mu$, 
while $h_\mu^S (k)$ is symmetric, fully diagonal and can be written as 
a derivative, namely~\footnote{We thank the referee for pointing this out
to us.}
\begin{equation}
   h_\mu^S(k) \;=\; 2 \kappa \frac{\partial}{\partial k_\mu}  
   \sin k_\mu \, B_0^{-\half}(k) 
   \label{eq:3.42}
\end{equation}
Consequently, $\int d^4 k \,D_1(k) = 0$ and we have 
\begin{equation}
     D_{0,0}\Bigl( U(a)\Bigr) \;=\; D_{0,0}(\,\identity\,) 
                             \,+\, \cO(a^2)
   \label{eq:3.48}           
\end{equation}
While relatively simple, this calculation fully illustrates the issues
encountered also at higher orders, where the number of terms to deal with 
grows rather quickly. 

\subsection{Tensor Component and the Second Order} 

The tensor part of $D_{0,0}$ first appears at second order in classical
lattice spacing. Using the technique described above we straightforwardly 
arrive (see also Appendix~\ref{app:secord}) at the expression in Fourier 
space, namely
\begin{equation}
   \frac{1}{\rho}\, D_2(k) \;=\; \sum_{\mu \nu}
   \biggl[\, \frac{1}{4} \frac{2r}{(4r-\rho)^3} 
            \frac{t_{\mu\nu}(k)}{B_0^{3/2}(k)} \,\biggr] \;
            \sigma_{\mu\nu} \times F_{\mu\nu}  \; + \; \ldots
   \label{eq:3.50}
\end{equation}
where $F_{\mu\nu}=i\,[A_\mu,A_\nu]$, 
$\sigma_{\mu\nu} = \frac{1}{2i}\,[\gamma_\mu,\gamma_\nu]$, and ``$\ldots$'' 
represents terms that are either odd in $k_\mu$ or are partial derivatives 
of analytic functions with respect to $k_\mu$, thus not contributing upon 
transition back from Fourier space. The expression in the bracket is 
a real--valued function of the momentum and it should be pointed out that 
from now on we use the symbol $B_0(k)$ to interchangeably denote a color--spin 
matrix proportional to identity, as well as its diagonal matrix element. 
The function $t_{\mu\nu}(k)$ is given by
\begin{equation}
  t_{\mu\nu}(k) \,=\, \sin^2 k_\mu \cos k_\nu \,+\,
                      \sin^2 k_\nu \cos k_\mu  \,-\,
                      \Bigl(\, 
                      4-\frac{\rho}{r} - \sum_\alpha \cos k_\alpha
                      \, \Bigr) \cos k_\mu \cos k_\nu
  \label{eq:3.52}  
\end{equation}
The constant $c^T=c^T(\rho,r)$ of Eq.~(\ref{eq:14.6}) is then specified by
\begin{equation}
   c^T(\rho,r) \;=\; 
   \frac{2r\rho}{(4r-\rho)^3} \,\frac{1}{(2\pi)^4} \,
            \int d^4 k \,\frac{t_{\mu\nu}(k)} {B_0^{3/2}(k)} 
   \label{eq:3.54}
\end{equation}
and doesn't depend on $\mu$,$\nu$ due to hypercubic symmetries. One can easily 
check that this result is identical to the one we obtained previously in
Ref.~\cite{KFL07A}.

The parameter dependence of $c^T$ is more transparent if one introduces a rescaled 
mass parameter
\begin{equation}
    \hrho \,\equiv \, \frac{\rho}{r}  \qquad\quad 
    \hrho \in (0,2)
   \label{eq:3.56}
\end{equation}
and the rescaled $B_0(k)$, namely
\begin{equation}
   z(k) \,\equiv \, \frac{1}{(2\kappa)^2} \,B_0(k) \,=\, 
                    (4r-\rho)^2 \, B_0(k)
   \label{eq:3.58} 
\end{equation}
With that we finally have
\begin{equation}
   c^T(\hrho,r) \;=\; 
     2 \hrho\, r^2 \,
            \int \frac{d^4 k}{(2\pi)^4} \,\frac{t_{\mu\nu}(k)} {z^{3/2}(k)} 
   \label{eq:3.60}
\end{equation}
where 
\begin{equation}
     z(k) \,=\, \sum_\mu \sin^2(k_\mu) \,+\, 
                r^2 \, (\,4-\hrho - \sum_\mu \cos k_\mu\,)^2
     \label{eq:3.62}
\end{equation}
depends on both $\hrho$ and $r$, while
\begin{equation}
  t_{\mu\nu}(k) \,=\, \sin^2 k_\mu \cos k_\nu \,+\,
                      \sin^2 k_\nu \cos k_\mu  \,-\,
                      (\, 
                      4-\hrho - \sum_\alpha \cos k_\alpha
                      \,) \cos k_\mu \cos k_\nu
  \label{eq:3.64}  
\end{equation}
only depends on $\hrho$.

\begin{figure}
\begin{center}
    \centerline{
    \hskip -0.00in
    \includegraphics[width=17.0truecm,angle=0]{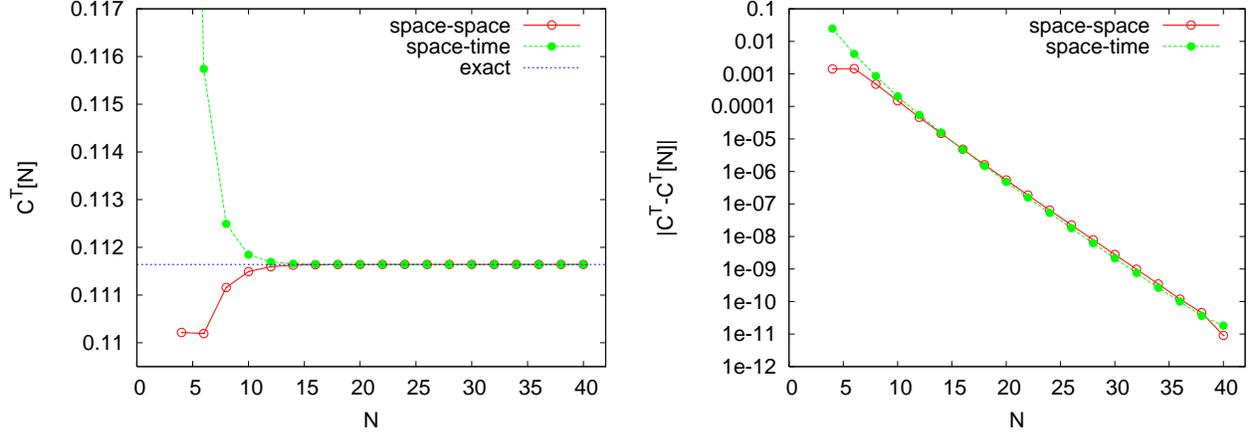}
     }
     \vskip -0.15in
     \caption{Riemann sums for proportionality constant $c^T$ of the tensor 
              term. These results were computed at $r=1$ and 
              $\kappa=0.19$ ($\rho=26/19$).}
     \vskip -0.4in 
     \label{fig:tensor}
\end{center}
\end{figure}

For explicit evaluation of $c^T$ we study the convergence of Riemann 
sums associated with the corresponding Brillouin zone integral.
More precisely, we partition the integration domain $[0,2\pi]^4$ into $N^4$ 
cubes of volume $(2\pi/N)^4 \equiv (\delta k)^4$ and define the corresponding 
Riemann sum $c^T[N]$ via
\begin{equation}
   \frac{1}{2 \hrho\, r^2} \, c^T[N] \,\equiv\, 
          \frac{1}{(2\pi)^4} \,\sum_k \, ( \delta k )^4 \,
          \frac{t_{\mu\nu}(k)} {z^{3/2}(k)} 
          \;=\; \frac{1}{N^4} \sum_k \, \frac{t_{\mu\nu}(k)} {z^{3/2}(k)}
   \label{eq:3.46}       
\end{equation}
where the discrete momenta associated with elementary cubes have components
\begin{equation}
   k_\mu \,=\, \cases{ 
            (\delta k) \, l_\mu \; & for $\mu=1,2,3$ \cr 
            (\delta k) \, (l_\mu+1/2) \; & for $\mu=4$ \cr}  
   \qquad\qquad
   l_\mu \,=\, 0,1,\ldots,N-1 
   \label{eq:3.46a}       
\end{equation}
Note that this choice of discrete momenta exactly corresponds to the situation
on the latticized torus of size $N$ with mixed periodic (spatial) and 
antiperiodic (time) boundary conditions, which is frequently a preferred setup 
in finite volume.~\footnote{The rationale for this is that it will allow us 
to argue that, for constant fields, our results on infinite lattice can be mapped 
exactly to this specific finite--volume case.} Due to the asymmetric choice 
of the discretization, the value of $c^T[N]$ at finite $N$ will depend on whether
the pair $\mu,\nu$ chosen for evaluation is of space--space or space--time kind, 
but the difference has to vanish in the $N\to\infty$ limit. 
In Fig.~\ref{fig:tensor} we show the dependence of $c^T[N]$ on $N$ for both 
cases, showing that they converge exponentially in $N$ to the common non--zero 
limit. The value labeled as ``exact'' includes only digits determined to be 
stabilized under increasing $N$ in the calculation.

\subsection{Scalar Component and the Fourth Order}
\label{sec:scalar}

Our main goal in this paper is the evaluation of classical limit for 
the scalar spinorial component of $D_{0,0}$, and its complete trace
in particular.
Since the expectation is that the leading term (up to a constant 
to be subtracted) will appear at the 4-th order, one has to carry out expansions 
described in Secs.~\ref{sec:expansion}, \ref{sec:first_order} up to that order.
Algebraic manipulations involved in this are rather extensive and we have used 
{\em Mathematica} for required symbolic manipulations. The result 
can be written in the form 
\begin{equation}
   {\Trs}_{sc}\, D_{0,0} ( U(a) )   \;=\;
   {\Trs}_{sc}\, D_{0,0} ( \identity ) \;+\; a^4\, I_4
                                       \;+\; \cO(a^6)
   \label{eq:3.66}  
\end{equation}
The odd powers are absent in the scalar spinorial component (do not contribute 
upon taking the spinorial trace of $D_{0,0}$). The quartic term is given by
\begin{equation}
   I_4 \,=\, \sum_\mu i_{4,\mu} \, {\Trs}_c\, A_\mu^4 \,+\,
   \sum_{\mu<\nu} \alpha_{\mu\nu} \, {\Trs}_c\, A_\mu A_\mu A_\nu A_\nu  \,+\,
   \sum_{\mu<\nu} \beta_{\mu\nu} \, {\Trs}_c\, A_\mu A_\nu A_\mu A_\nu
   \label{eq:3.80}  
\end{equation}

To describe the results of calculation for constants 
$i_{4,\mu}$, $\alpha_{\mu\nu}$, $\beta_{\mu\nu}$ we now have to introduce few 
conventions. All the coefficients to compute will be defined by functions in 
the momentum space of the form
\begin{equation}
   \Fhat (k;j,\chat) \,=\, \frac{1}{B_0(k)^{j}} \,\sum_{n}\, \chat(n)
   \, \cos(k_1)^{n_1} \cos(k_2)^{n_2} \cos(k_3)^{n_3} \cos(k_4)^{n_4}
   \label{eq:3.70}  
\end{equation}
where $n\equiv(n_1,n_2,n_3,n_4)$ is a four--component vector of non--negative
integers, $j$ is a single half--integer and $\chat (n)$ is real--valued.
Each function of interest is thus specified by $j$ and the finite list of 
non--zero coefficients $\chat (n)$.  

To reduce the amount of information needed for evaluation of the coefficients, 
we will use the fact that, by means of the inverse Fourier transform, it is 
only the mean value of ${\hat F}$ that is relevant. More precisely, we are 
interested only in
\begin{equation}
    F(j,\chat) \,\equiv\, \int \frac{d^4 k}{(2\pi)^4} \,{\hat F}(k;j,\chat) 
           \,=\, \sum_n \chat (n) \, I(j,n)
   \label{eq:3.72}  
\end{equation}
where we introduced the notation $I(j,n)$ for integrals of the form
\begin{equation}
    I(j,n) \,\equiv\, \int \frac{d^4 k}{(2\pi)^4} \,  
    \frac{\cos(k_1)^{n_1} \cos(k_2)^{n_2} \cos(k_3)^{n_3} \cos(k_4)^{n_4}}
         {B_0(k)^j}
   \label{eq:3.74}  
\end{equation}
Since $I(j,n)$ is completely symmetric with respect to indices $n_\mu$,
we can restrict the sum over $n$ in (\ref{eq:3.72}) only to $n$ such
that $n_1 \ge n_2 \ge n_3 \ge n_4$, and redefine the coefficients 
$\chat (n) \rightarrow \,c(n)$ accordingly. We then have
\begin{equation}
    F(j,c) \,\equiv\, \sum_{n_1 \ge n_2 \ge n_3 \ge n_4} c(n) \, I(j,n)
   \label{eq:3.76}  
\end{equation}
and we specify our results by listing $j$ and $c(n)$ for each case in 
question.

The last piece of convention we need to specify concerns the choice
of discretization for performing Riemann sums in the actual evaluation of 
momentum integrals. While final results have to be independent of that 
choice, our discussion of classical limits in the finite volume 
(see Sec.~\ref{sec:finite_volume}) can be mapped directly on the discussion 
here if Riemann sums are defined in a particular manner. Moreover, 
consistency over different discretizations
represents an additional check on our results. In what follows, we will
refer to the discretization defined by Eq.~(\ref{eq:3.46a}) as
{\em antiperiodic}, while the discretization where
\begin{equation}
   k_\mu \,=\, (\delta k) \, l_\mu 
   \qquad\quad \mu=1,2,3,4  
   \qquad\qquad l_\mu \,=\, 0,1,\ldots,N-1 
   \label{eq:3.78}  
\end{equation}
will be referred to as {\em periodic}. We emphasize that when we speak of
periodic or antiperiodic case, we also imply that the corresponding reduction 
$\chat (n) \rightarrow c(n)$ has been performed in a manner that is consistent 
with the remnant symmetries of the Riemann sum used. More precisely, for 
the periodic discretization the Riemann sums $I(j,n)[N]$ are completely symmetric 
with respect to indices $n_\mu$ and the reduction is thus as described in 
previous paragraph. 
However, for the antiperiodic case the Riemann sum is only symmetric with
respect to exchange of spatial components $n_1$, $n_2$ and $n_3$, and one thus 
has to keep track of larger set of coefficients $c(n)$. It is only with this
prescription that the exact correspondence between the finite and infinite volume
situations is realized in case of antiperiodic discretization of the integral. 
One implication of this is that, for antiperiodic case at finite $N$, 
the time--like components of vectors and tensors 
will not be exactly equal to space--like components. However, the difference 
(and recovery of full hypercubic symmetries of the infinite lattice) has to take 
place in the $N\to\infty$ limit. In what follows, we will show results for 
both types of discretization. Perfect agreement has always been found as expected, 
and we thus provide the list of reduced coefficients $c(n)$ of the periodic 
case only, which is more economic.
\begin{figure}
\begin{center}
    \centerline{
    \hskip -0.00in
    \includegraphics[width=17.0truecm,angle=0]{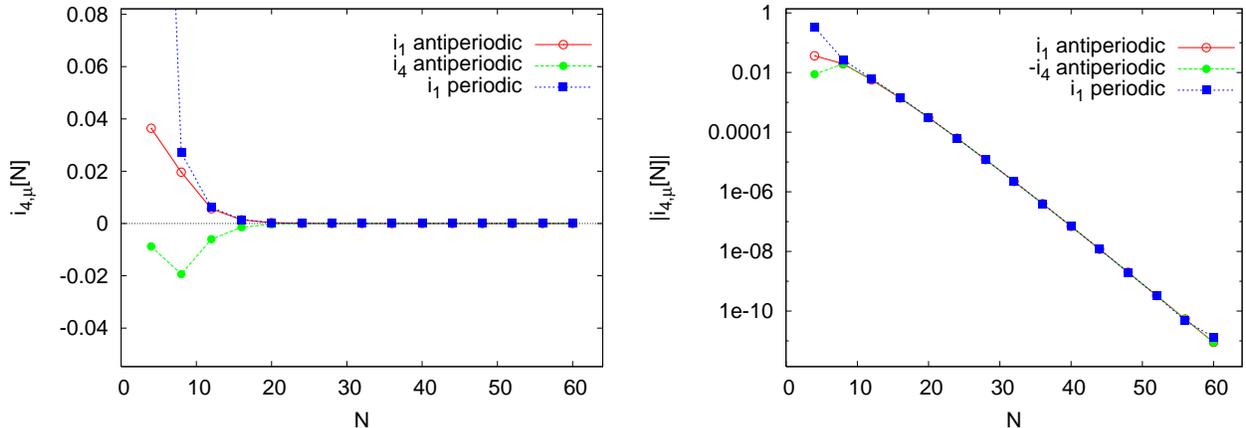}
     }
     \vskip -0.15in
     \caption{Riemann sums for coefficients $i_{4,\mu}$ of Eq.~(\ref{eq:3.80}). 
              These results were computed at $r=1$ and $\kappa=0.19$ 
              ($\rho=26/19$).}
     \vskip -0.4in 
     \label{fig:i_4}
\end{center}
\end{figure}

\subsubsection{The Results}
\label{sec:scalar_quart}

The first group of terms in Eq.~(\ref{eq:3.80}) is not gauge invariant and so 
constants $i_{4,\mu}$ are expected to be zero. We have numerically evaluated these 
constants using the result for corresponding coefficients $c(n)$ obtained from 
{\em Mathematica}. Fig.~\ref{fig:i_4} shows the typical convergence of the 
associated Riemann sums $i_{4,\mu}[N]$ both in antiperiodic and periodic 
discretizations. For antiperiodic case both the time--like and the space--like 
case is shown to see that they are different at finite $N$. For $N \to \infty$ 
all three sequences are expected to approach the common limit which is apparently 
zero in this case. The convergence is again exponential in $N$.

The second and the third group of terms contain combinations that appear in
($\mu$, $\nu$ fixed) 
\begin{equation}
  {\Trs}_c F_{\mu\nu} F_{\mu\nu} \,=\, -{\Trs}_c \,[A_\mu,A_\nu][A_\mu,A_\nu] 
  \,=\, 2 \,{\Trs}_c \,(A_\mu A_\mu A_\nu A_\nu - A_\mu A_\nu A_\mu A_\nu)
  \label{eq:3.82}   
\end{equation}
and thus, in light of Conjecture~1, we anticipate that 
$\beta_{\mu\nu} = -\alpha_{\mu\nu}$. Also, due to hypercubic symmetries 
we must have that $\alpha_{\mu\nu}\equiv \alpha$ and $\beta_{\mu\nu}\equiv \beta$
are independent of $\mu\ne\nu$. Since our results were obtained 
using a symbolic algebra software, verifying these relations is not only 
a consistency check for the expected result, but also an internal check
of our programs. 

\begin{figure}
\begin{center}
    \centerline{
    \hskip -0.00in
    \includegraphics[width=13.5truecm,angle=0]{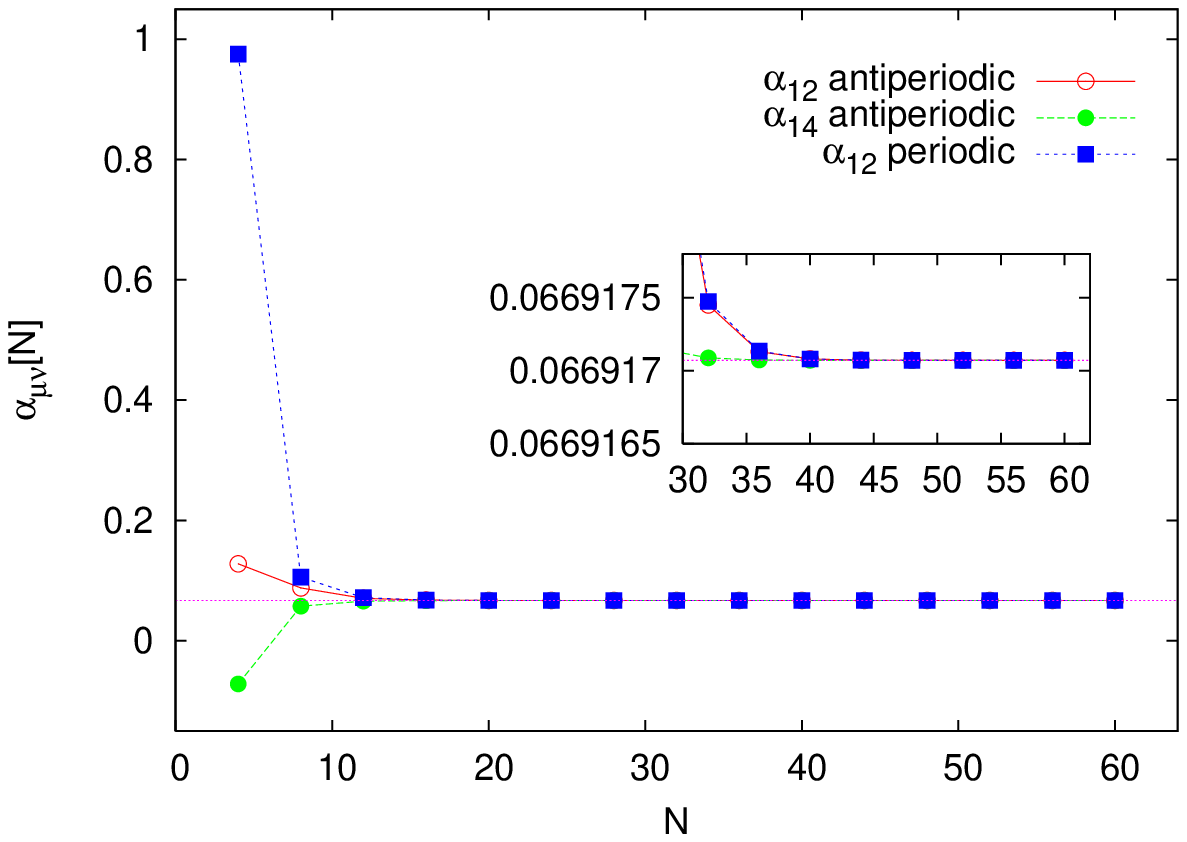}
     }
    \hskip 0.2in
    \centerline{
    \hskip -0.00in
    \includegraphics[width=13.5truecm,angle=0]{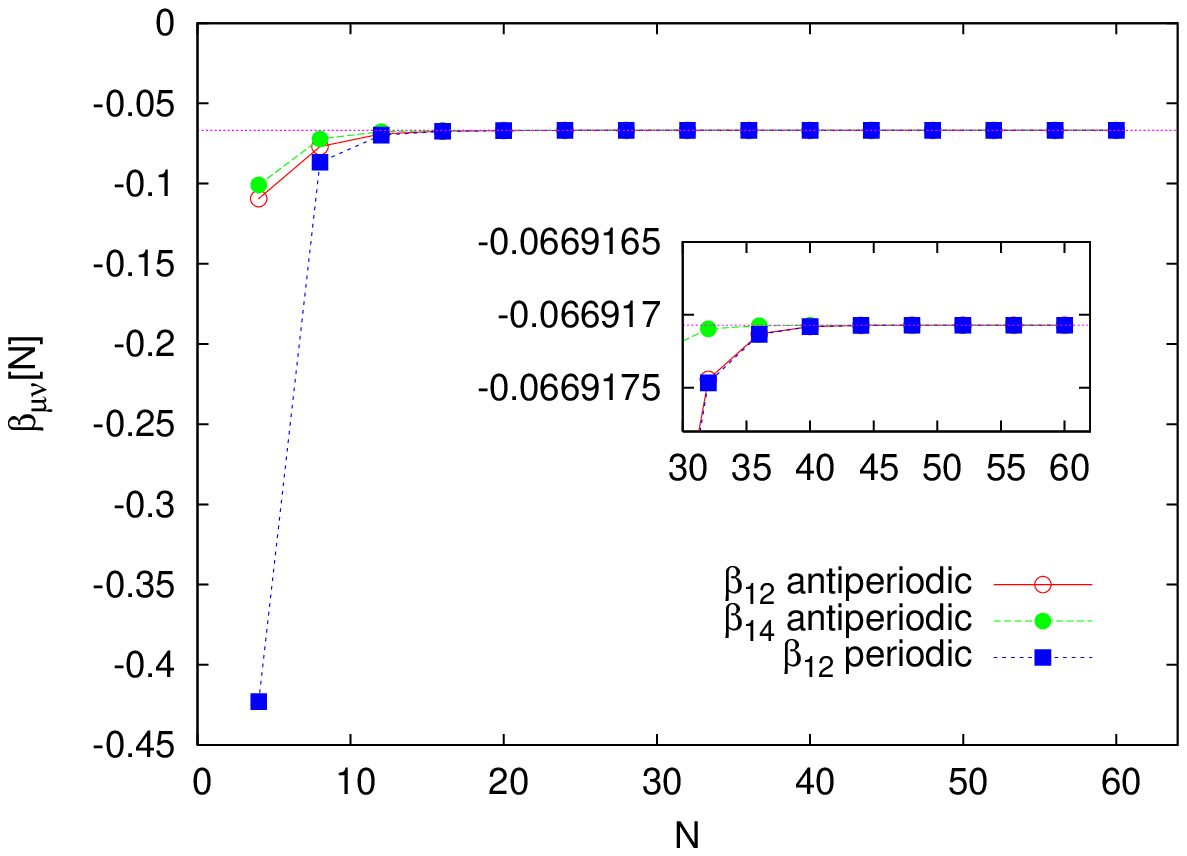}
     }
     \vskip -0.00in
     \caption{Riemann sums for constants $\alpha_{\mu\nu}$ (top) and 
     $\beta_{\mu\nu}$ (bottom) of the scalar term (see Eq.~(\ref{eq:3.80})). 
     These results were computed at $r=1$ and $\kappa=0.19$ ($\rho=26/19$).}
     \vskip -0.1in 
     \label{fig:alpha_beta}
\end{center}
\end{figure}

The set of reduced coefficients specifying constants $\alpha_{\mu\nu}$ and
$\beta_{\mu\nu}$ are given in Tables 1 and 2 respectively ($j=9/2$).  
In Fig.~\ref{fig:alpha_beta} (top) we show the typical convergence of Riemann 
sums for $\alpha_{\mu\nu}$ in both periodic and antiperiodic discretizations of 
the integral. For periodic case only $\alpha_{12}$ is shown since full hypercubic 
symmetry in this case guarantees that, even at finite $N$, $\alpha_{\mu\nu}[N]$ 
is independent
of $\mu$, $\nu$. Our results, obtained using symbolic algebra software, indeed 
comply with this. In the antiperiodic case we show both the representative of 
space--space combination ($\alpha_{12}$) and the representative of space--time 
combination ($\alpha_{14}$). All other possibilities for space--space and 
space--time combinations are exactly the same as the ones shown (even at finite $N$)  
as expected from the restricted hypercubic symmetry. The convergence of the three 
Riemann sums is exponential, as was in all the previous cases, and the insert shows 
that they approach a common non--zero limit whose value $\alpha$ can be easily 
extracted. The analogous results in case of $\beta_{\mu\nu}$ are shown in the bottom
plot of Fig.~\ref{fig:alpha_beta}, and one can immediately see that, indeed,
$\beta=-\alpha$. 

Combining all the results for the scalar spinorial component of $D_{0,0}$ together,
we have from Eq.~(\ref{eq:3.66}) that
\begin{equation}
   {\Trs}_{sc}\, D_{0,0} ( U(a) )   \;=\;
   {\Trs}_{sc}\, D_{0,0} ( \identity ) \;+\;  
   a^4\, \frac{\alpha}{4} \,\sum_{\mu\nu} {\Trs}_c \, F_{\mu\nu}F_{\mu\nu}
                 \;+\; \cO(a^6)
   \label{eq:3.84}  
\end{equation}
implying that Conjecture 1 (see Eq.(\ref{eq:14.5})) is indeed true for the class
of constant fields. The associated proportionality constant is given by
\begin{equation}
   c^S(\rho,r) \,=\, \frac{\alpha(\rho,r)}{4}
   \qquad\longrightarrow\qquad
   c^S\Bigl(\,\frac{26}{19},1\,\Bigr) = 0.01672926781
   \label{eq:3.86}  
\end{equation}
where the quoted value at $\rho=26/19$ and $r=1$ (only valid digits are given)
was extracted from the data shown in Fig.~\ref{fig:alpha_beta}.


\section{Finite Volume and Non--Constant Fields}
\label{sec:finite_volume}

As already mentioned in the previous section, we have analyzed the case 
of constant fields in the infinite volume in a way that can be directly mapped
on to the situation in finite volume with specific boundary conditions. 
We now discuss this correspondence explicitly. Consider a 4--d hypercube in 
the continuum with side $L_p$ (in physical units). Superimposing a hypercubic 
lattice with $N^4$ sites implies the classical lattice spacing $a \equiv a_N$ 
such that
\begin{equation}
   L_p \,=\, N \,a_N
   \label{eq:4.10}
\end{equation}
and the classical continuum limit is achieved as $N \to \infty\,$ ($a_N \to 0$).
Proper specification of the operator now involves also a choice of boundary 
conditions, and we will consider (see below) two standard cases that admit 
a diagonal Fourier--space representation for constant fields. In particular,
we will refer to the situation with periodic boundary conditions in all directions
as ``periodic bc'', and to the case with periodic boundary conditions
in spatial ($\mu=1,2,3$) directions and antiperiodic in time ($\mu=4$) 
as ``antiperiodic bc''. The formulas (\ref{eq:3.4}) relating the
direct and Fourier representations carry over to these finite volume cases
except that the infinite sum is replaced by the finite sum, and the integral 
of the inverse transform is replaced by
\begin{equation}
  O_{n,m} \,=\, \frac{1}{(2\pi)^4} \int d^4 k \,e^{i(n-m)k} \,O(k)
  \qquad \longrightarrow \qquad
  O_{n,m} \,=\, \frac{1}{N^4} \sum_k \,e^{i(n-m)k} \,O(k)
  \label{eq:4.12}
\end{equation}
with sum only running over the associated $N^4$ discrete momenta. 
\smallskip

\noindent
The relation to the infinite--volume case is based on the following 
considerations.
\smallskip

\noindent {\em (i)} With periodic bc, the $N^4$--dimensional Fourier space is 
spanned by plane waves with momenta given by Eq.~(\ref{eq:3.78}), namely those 
involved in the periodic discretization (into $N^4$ pieces) of the Brillouin zone 
discussed in case of infinite volume. Similarly, for antiperiodic bc 
the associated momenta are given by Eq.~(\ref{eq:3.46a}) of the antiperiodic 
discretization. 
\smallskip

\noindent{\em (ii)} The form of Fourier representation $D(k)$ of the overlap 
operator for the above finite volume setups (and constant fields) is identical 
to the infinite volume case but restricted to the corresponding discrete momenta. 
\smallskip

\noindent{\em (iii)} Using {\em (i), (ii)} and comparing (\ref{eq:4.12}) with the
rule for evaluating Riemann sums in case of infinite volume 
(see e.g. Eq.~(\ref{eq:3.46})) we can immediately see the following correspondence
\begin{equation}
     D_{0,0}^{fin}(a,N) \,\equiv \, D_{0,0}^{inf}(a)[N] 
     \label{eq:4.14}
\end{equation}
where the superscripts {\em ``inf''} and {\em ``fin''} refer to infinite 
and finite volume cases respectively. In other words, the matrix 
$D^{fin}_{0,0}(a,N)$ (for finite volume setups discussed here) is equal to 
the corresponding Riemann sum (with $N^4$ terms) for the inverse Fourier transform 
representation of $D^{inf}_{0,0}(a)$. The same constant background is of course
implicitly assumed in both cases.
\smallskip

\noindent{\em (iv)} Riemann sums in question converge exponentially in $N$, 
as emphasized in previous sections, and we can conclude from (\ref{eq:4.14})
that for large $N$
\begin{equation}
     D_{0,0}^{fin} (a,N) \,=\, D_{0,0}^{inf}(a) \,+\,
     \cO \Bigl(\exp (-\,C(a)\,N) \Bigr) 
     \label{eq:4.16}
\end{equation}
with strictly positive $C(a)$ (also at $a=0$).

From the last equation it follows that, in the case of constant background 
fields,
the Taylor expansion (in $a$) of $D_{0,0}$ in infinite volume supplies an 
asymptotic expansion for the case of finite volume (with corections 
behaving as $\exp(-C(a) L_p/a)$. Moreover, considering the continuum limit 
at fixed $L_p$ (so that $a_N\equiv L_p/N$) we obtain that
\begin{equation}
   \lim_{N\to\infty} \,
   \frac{\Trs_{sc} \Bigl( D^{fin}_{0,0}(a_N,N) \,-\, 
         D^{fin}_{0,0}(0,N) \Bigr)}
        {a_N^4\, \Trs_c F_{\mu\nu} F_{\mu\nu}}
   \;=\;
   \lim_{N\to\infty} \,
   \frac{\Trs_{sc} \Bigl( D^{inf}_{0,0}(a_N) \,-\, D^{inf}_{0,0}(0) \Bigr)}
        {a_N^4\, \Trs_c F_{\mu\nu} F_{\mu\nu}}
   \;\equiv\; c^S 
   \label{eq:4.18}
\end{equation} 
where $c^S$ is the constant that we computed in the infinite volume. The above 
equation expresses the fact that the scalar component of $D_{0,0}$ has 
the same classical limit in both finite and infinite volumes, as required 
and expected on the basis of locality. Analogous result obviously holds also 
for the tensor component.

\subsection{Non-Constant Fields}
\label{sec:non-constant}

Apart from the fact that the finite-volume setups are actually those relevant for
practical lattice QCD calculations, their advantage for the present discussion 
is that one can evaluate the associated classical limit numerically. In other words,
one can obtain the left-hand side of Eq.~(\ref{eq:4.18}) via direct numerical 
evaluation of $D^{fin}_{0,0}$ for {\em arbitrary} classical background. Performing 
a sequence of such calculations with increasing $N$ then allows us to infer the 
$N\to\infty$ ($a_N \to 0$) limit of the ratio. For the case of constant fields, 
such calculations indeed exactly reproduce our results obtained via the expansion 
in classical lattice spacing. The utility of this approach however mainly lies in 
the fact that it allows us to evaluate classical limits also for non--constant 
backgrounds which were not included in our considerations up to this point.
 
To perform such a calculation, we set $L_p=1$ and use arbitrarily selected classical 
backgrounds. One of the cases that we studied is specified by
\begin{equation}
  A_\mu(t,x,y,z) = \left(1 + {\sin 2\pi t \,\cos 2\pi x \mu \over
  12\pi}\right) a_\mu + \left(1 + {\cos 2\pi z\over
  2\pi}\right) a_{3-\mu}
  \label{eq:4.20}
\end{equation}
where we use $\mu \in \{0,1,2,3\}$ and the coordinate--labeling correspondence 
$0\leftrightarrow t$, $1 \leftrightarrow x$, $2\leftrightarrow y$, 
$3\leftrightarrow z$ for convenience. The constant field $a_\mu$ is specified 
in Appendix~\ref{app:constant}.

\begin{figure}
\begin{center}
    \centerline{
    \hskip -0.00in
    \includegraphics[width=13.5truecm,angle=0]{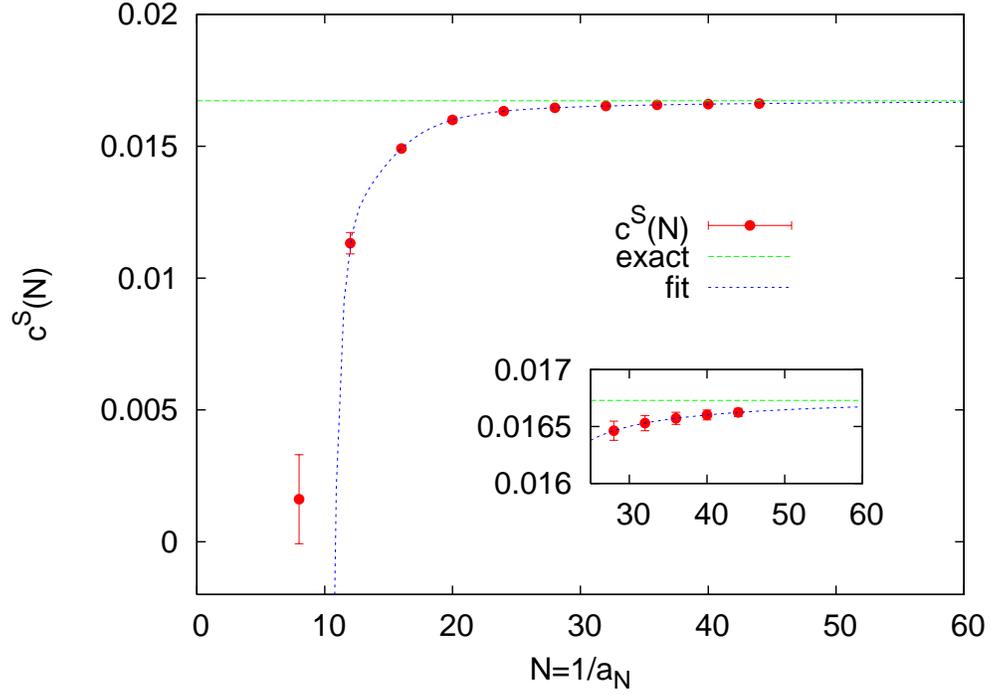}
     }
    \hskip 0.2in
    \centerline{
    \hskip -0.00in
    \includegraphics[width=13.5truecm,angle=0]{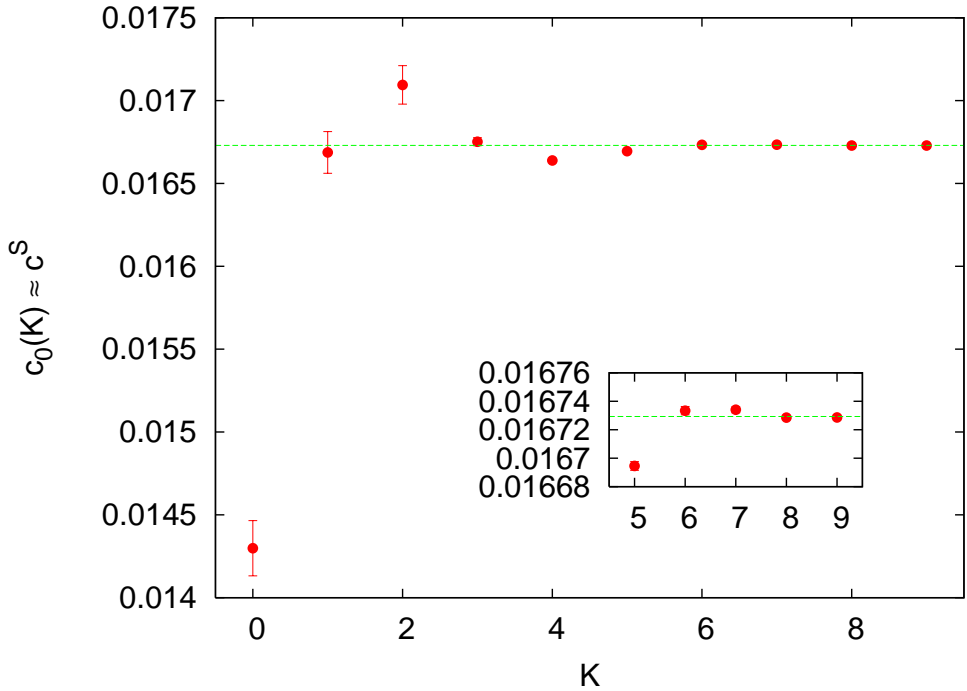}
     }
     \vskip -0.00in
     \caption{Evaluation of classical limit on a finite torus ($L_p=1$)
     for non-constant background field (\ref{eq:4.20}). The meaning of the data 
     is explained in the text. The overlap operator used is specified by $r=1$ 
     and $\kappa=0.19$ ($\rho=26/19$).}
     \vskip -0.1in 
     \label{fig:cs_finite_N}
\end{center}
\end{figure} 

For given $N$, we evaluated the ratio $c^S(N=1/a_N)$ at 10 randomly selected points 
on the unit torus. The same points, specified in Appendix~\ref{app:constant}, were
used for each $N$. In Fig.~\ref{fig:cs_finite_N} (top) we plot the mean value of 
$c^S(N)$ over this sample with ``error bars'' representing the square root of 
the associated variance. The horizontal line shows the value of $c^S$ obtained via 
expansion in classical lattice spacing in infinite volume (\ref{eq:3.86}). 
As can be seen quite clearly, the mean approaches this predicted value for large $N$ 
with variance shrinking toward zero at the same time. To guide the eye, we also 
included a fit to $N=1/a_N$ dependence of $c^S$ in the form
\begin{equation}
    c^S(N) \,=\, \sum_{k=0}^{K} c_k \, \biggl( \frac{1}{N} \biggr)^{2k} 
    \label{eq:4.22}
\end{equation}
Note that this form is motivated by the fact that the odd powers of lattice
spacing are not present in the expansion of $D_{0,0}$ in infinite volume.      
The above fitting form is neglecting the terms exponentially small in $N$, which 
are present, but their relative contribution decays very fast with increasing $N$. 
The standard fit shown in Fig.~\ref{fig:cs_finite_N} uses all points with 
$N \ge 12$ (treating the associated rooted variances as error bars) and $K=6$. 
The coefficient $c_0$ of the fit represents the estimate of $c^S$, 
and the obtained value agrees with that of Eq.~(\ref{eq:3.86}) to relative 
precision of about two parts in $10^4$.

To further check the robustness of the agreement found above (universality of 
the classical limit for constant and non--constant configurations), let us now 
examine a different procedure for estimating $c^S$ on background (\ref{eq:4.20}). 
For any given point on the torus, consider evaluating the ratio (\ref{eq:4.18}) over 
some range of $N$ (such as one shown in Fig.~\ref{fig:cs_finite_N} (top)), and then 
fitting this dependence to the form (\ref{eq:4.22}) with progressively increasing 
value of $K$.\footnote{Note that all the points in the chosen range contribute with 
equal weight in such fit -- there are no errorbars. Indeed, classical limit is not 
a ``statistical'' notion and has to be consistently reproduced at all non--singular
points and for all classical configurations.} 
For each $K$, the coefficient $c_0 \equiv c_0(K)$ of the fit represents the estimate 
of $c^S$. Since the expansion (\ref{eq:4.22}) is asymptotically valid at large
values of $N$, the accuracy of such estimates will depend on the fitting range
chosen. In order to be very conservative, we use the whole range displayed 
in Fig.~\ref{fig:cs_finite_N} (top), i.e. all the points with $8\le N \le 44$. 
In Fig.~\ref{fig:cs_finite_N} (bottom) we show the average of $c_0(K)$ so obtained
over the same sample of random points that we used above. The ``errorbars''
on this plot are the square roots of variance over this sample, and are decaying
rapidly with increasing $K$. This is expected since the classical limit has to
be reproduced at each point individually and the estimates are expected to get better
with increasing $K$. The horizontal line on the plot represents the value of $c^S$ 
quoted in Eq.~(\ref{eq:3.86}). As one can see quite clearly, $c_0(K)$ settles 
approximately at this expected value. In fact, for the largest value of $K$ ($K=9$) 
the relative difference between $c^S$ and $c_0$ is about three parts in $10^5$.

\section{Values of Related Constants}
\label{sec:values}

Given the results presented in Secs.~\ref{sec:const_fields}, \ref{sec:finite_volume} 
there is little doubt that the universal classical limit for the scalar spinorial 
part of $D_{0,0}$ exists in the case of overlap operator, and that it is proportional 
to $F_{\mu\nu}(0) F_{\mu\nu}(0)$~\cite{Hor06B}. Indeed, we obtained fully consistent 
results showing this for finite and infinite volumes, for constant and non--constant 
fields, and for different boundary conditions in finite volume. Since matrix elements 
of $D$ are local functions of the gauge field~\cite{ov_loc}, the lattice operator
\begin{equation}
   O(n,U) \equiv \frac{1}{c^S} \,
                     {\Trs}_{sc}\, \Bigl(\, D_{n,n} (U) \,-\, 
                          D_{n,n} ( \identity )\, \Bigr) 
   \label{eq:5.2}
\end{equation} 
represents a valid definition of ${\Trs}_c F_{\mu\nu} F_{\mu\nu}$. For certain 
applications, such as those arising in the studies of QCD vacuum structure, it is 
necessary to deal with properly 
normalized operators in which case it is important that sufficiently precise values 
of $c^S$ and free-field subtraction constants are available. In this section, we 
provide some of this information both for the scalar and tensor cases.

As is well known, the properties of the overlap Dirac operator, such as its range
of locality~\cite{ov_loc}, are quite sensitive to the value of the negative mass
parameter $\rho$. Since various choices are currently being used in practice,
we computed $c^S(\rho,r)$ and $c^T(\rho,r)$ for the range of $\rho$--values
and the most commonly used case of $r=1$. To obtain the results listed 
in Table~\ref{tab:c_values}, we used the expansion method described in 
Sec.~\ref{sec:const_fields}.
The last column lists the maximal discretization parameter $N$ used in evaluating
the associated Riemann sums. Note that one needs to use finer discretization
(larger $N$) close to boundary values $\rho=0$ and $\rho=2$ to achieve given 
precision.

\setcounter{table}{2}

\begin{table}[t]
  \centering
  \begin{tabular}{ccccc}
  \hline\\[-0.4cm]
  \multicolumn{1}{c}{$\quad\rho\quad$}  &
  \multicolumn{1}{c}{$\qquad\kappa\qquad$}  &
  \multicolumn{1}{c}{$\qquad\qquad c^S/\rho\qquad\qquad$}  &
  \multicolumn{1}{c}{$c^T/\rho$} &
  \multicolumn{1}{c}{$\quad$ max $N \quad$} \\[2pt]
  \hline\\[-0.4cm]
  0.2 & 0.131579 & $7.2764827\times 10^{-3}$ & $2.30823770\times 10^{-2}$ & 160\\
  0.3 & 0.135135 & $7.2792339\times 10^{-3}$ & $2.78332999\times 10^{-2}$ & 120\\
  0.4 & 0.138889 & $7.3198516\times 10^{-3}$ & $3.26397580\times 10^{-2}$ & 100\\
  0.5 & 0.142857 & $7.4053234\times 10^{-3}$ & $3.75015319\times 10^{-2}$ & 100\\
  0.6 & 0.147059 & $7.5441944\times 10^{-3}$ & $4.24178729\times 10^{-2}$ & 80\\
  0.7 & 0.151515 & $7.7470040\times 10^{-3}$ & $4.73873153\times 10^{-2}$ & 80\\
  0.8 & 0.156250 & $8.0268757\times 10^{-3}$ & $5.24074212\times 10^{-2}$ & 80\\
  0.9 & 0.161290 & $8.4003244\times 10^{-3}$ & $5.74744291\times 10^{-2}$ & 80\\
  1.0 & 0.166667 & $8.8883824\times 10^{-3}$ & $6.25827634\times 10^{-2}$ & 80\\
  1.1 & 0.172414 & $9.5181966\times 10^{-3}$ & $6.77243431\times 10^{-2}$ & 80\\
  1.2 & 0.178571 & $1.0325346\times 10^{-2}$ & $7.28875894\times 10^{-2}$ & 80\\
  1.3 & 0.185185 & $1.1357281\times 10^{-2}$ & $7.80559770\times 10^{-2}$ & 80\\
  1.4 & 0.192308 & $1.2678582\times 10^{-2}$ & $8.32058759\times 10^{-2}$ & 80\\
  1.5 & 0.200000 & $1.4379234\times 10^{-2}$ & $8.83032584\times 10^{-2}$ & 100\\
  1.6 & 0.208333 & $1.6588164\times 10^{-2}$ & $9.32985300\times 10^{-2}$ & 120\\
  1.7 & 0.217391 & $1.9496369\times 10^{-2}$ & $9.81181443\times 10^{-2}$ & 120\\
  1.8 & 0.227273 & $2.3398534\times 10^{-2}$ & $1.02650441\times 10^{-1}$ & 180\\
\hline 
\end{tabular}
\caption{The proportionality constants $c^S$ and $c^T$ for various values 
         of $\rho$ at $r=1$.}
\label{tab:c_values} 
\end{table}

\begin{figure}
\begin{center}
    \centerline{
    \hskip -0.00in
    \includegraphics[width=13.5truecm,angle=0]{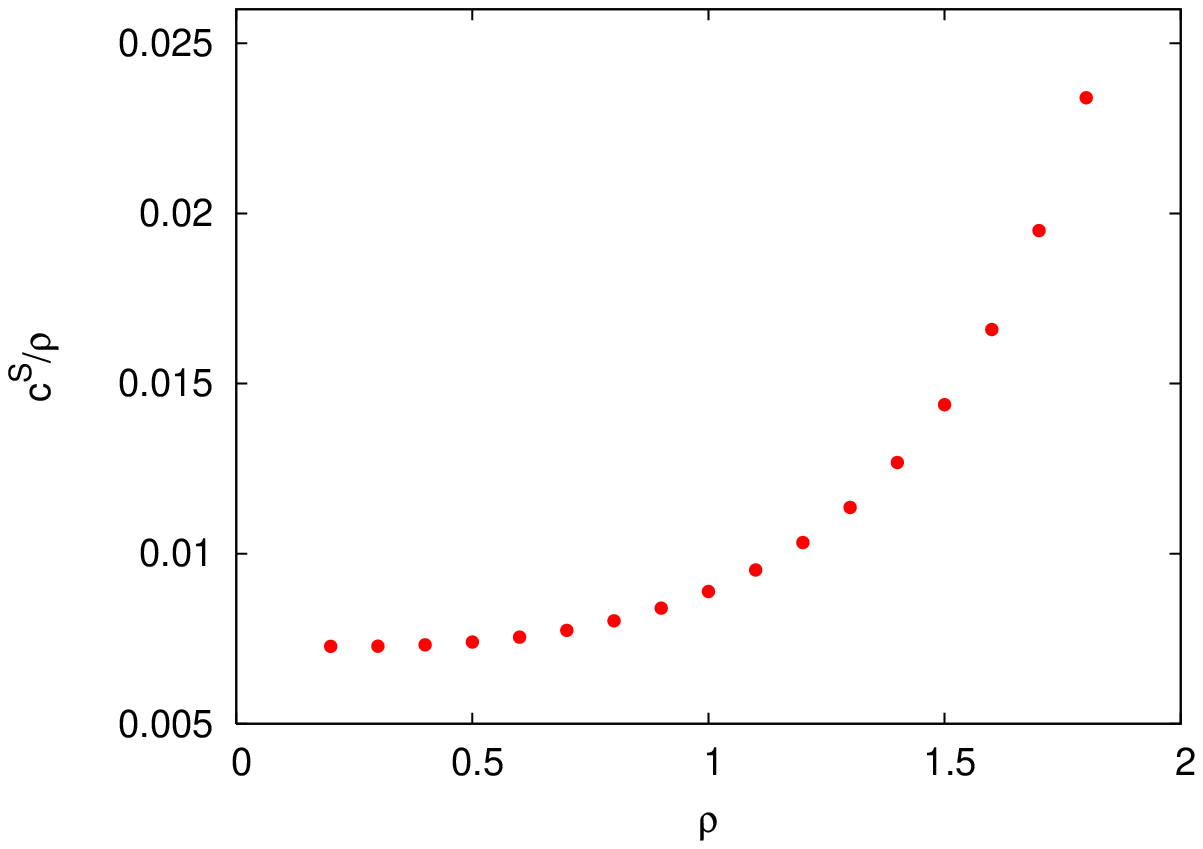}
     }
    \hskip 0.2in
    \centerline{
    \hskip -0.00in
    \includegraphics[width=13.5truecm,angle=0]{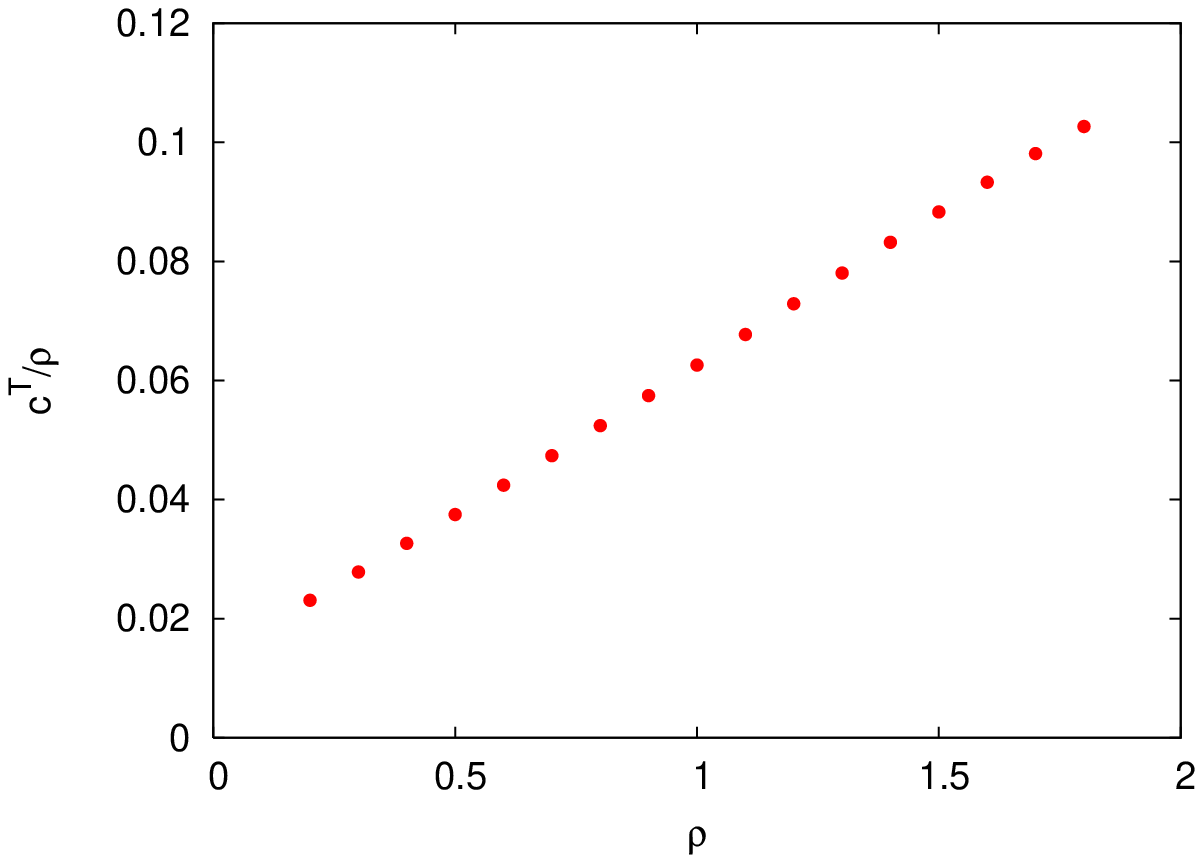}
     }
     \vskip -0.00in
     \caption{Dependence of rescaled constants $c^S$ (top) and $c^T$
     (bottom) on the negative mass parameter $\rho$ at $r=1$.}
     \vskip -0.1in 
     \label{fig:rho_dependence}
\end{center}
\end{figure}

In Figure~\ref{fig:rho_dependence}, we plot the $\rho$--dependence of both 
$c^S/\rho$ and $c^T/\rho$. It is worth pointing out that in the tensor case 
the behavior appears to be almost exactly linear in $\rho$. However, closer 
inspection suggests that the deviations from exact linearity are in fact larger 
than the estimated errors of calculated values. Nevertheless, the purely linear 
approximation is very precise in the range of $\rho$-values studied.

Finally, we have computed the subtraction constants 
${\Trs}_{sc}\,D_{0,0} (\identity)$
for various sizes of a symmetric lattice with both periodic and antiperiodic
boundary conditions. The results, summarized in Table~\ref{tab:sub_values}, 
were computed at $\rho=26/19$ and $r=1$. These are the values of the parameters 
used by the Kentucky group in the studies of QCD vacuum structure 
(see e.g. \cite{Hor05B,Hor03A,Hor05A}) as well as in the studies of hadron 
spectroscopy (see e.g. \cite{Mat04A,Mat07A}). In the space--time symmetric 
geometry considered here, the utility of these results is mainly for the former.
To obtain the constants quoted in Table~\ref{tab:sub_values}, we directly 
evaluated ${\Trs}_{sc}\,D_{0,0} (\identity)$ from the finite spectral sum over 
explicitly available Fourier modes. The calculation was done in double precision 
and we estimate that after rounding errors, these values are good to at least 
thirteen digits. Note that the convergence to the infinite volume value is very
fast and that at $N=40$ the periodic and the antiperiodic cases already agree
to 13 digits.

\begin{table}[t]
  \centering
  \begin{tabular}{ccc}
  \hline\\[-0.4cm]
  \multicolumn{1}{c}{$\quad N \quad$}  &
  \multicolumn{1}{c}{$\qquad$ periodic bc $\qquad$}  &
  \multicolumn{1}{c}{$\qquad$ antiperiodic bc $\qquad$} \\[2pt]
  \hline\\[-0.4cm]
  04 &  21.4358235894802  &  21.4010898463124\\
  08 &  21.3940949198520  &  21.3935062781772\\
  12 &  21.3931417096880  &  21.3931106079229\\
  16 &  21.3930879676987  &  21.3930856400191\\
  20 &  21.3930838480598  &  21.3930836424736\\
  24 &  21.3930834815890  &  21.3930834616044\\
  28 &  21.3930834459514  &  21.3930834438807\\
  32 &  21.3930834422695  &  21.3930834420471\\
  36 &  21.3930834418726  &  21.3930834418505\\
  40 &  21.3930834418356  &  21.3930834418306\\
\hline 
\end{tabular}
\caption{The subtraction constants ${\Trs}_{sc}\,D_{0,0} (\identity)$
         on $N^4$ lattice with periodic and antiperiodic (in time)
         boundary conditions. These results were calculated at 
         $\rho=26/19$ and $r=1$.}
\label{tab:sub_values} 
\end{table}

\section{Summary}

We have used analytic and numerical techniques to evaluate the classical 
limit of the gauge operator ${\Trs}_{sc}\,D_{0,0}(U)$ with 
$D\equiv D^{(\rho,r)}$ 
being the overlap Dirac matrix.~\footnote{All the results presented in this 
manuscript were obtained in the Summer of 2006 and were partially discussed
by the second author in Lattice 2006 talk that focused on coherent LQCD. 
In the Summer of 2007 we were informed by David Adams~\cite{Ada07} that 
he has just computed the constant $c^S(\rho,r)$ as well. His approach leads 
to an integral expression with integrand different from ours, but yielding 
identical final answers after integration. We hope that his derivation will 
be publicly available in due course.} 
As suggested on general grounds~\cite{Hor06B}, 
we found that the limit is proportional to 
${\Trs}_c \, F_{\mu\nu}(0) F_{\mu\nu}(0)$ 
of the associated classical background after the subtraction of 
the free--field constant. 
Accordingly, the version of coherent LQCD where ${\Trb}\,D(U)$ serves as 
a basis for the gauge action~\cite{Hor06B} represents a valid regularization
of QCD. In addition, this operator is expected to be useful as 
a natural partner to the overlap--based definition of pseudoscalar (topological) 
density both in studies of QCD vacuum structure and in standard applications
of hadronic physics (such as calculations of glueball masses). In the 
former case it is of interest to explore the relation between the fundamental 
structure seen in topological 
density~\cite{Hor02D,Hor03A,Hor05A,Ale05A,Ahm05A,Ilg07A}, and 
the structure in scalar density that is expected to be visible if 
the overlap--based definition (\ref{eq:5.2}) is used.

Techniques used here for the scalar case were applied in parallel also 
to the tensor component of $D_{0,0}$. In our previous work~\cite{KFL07A} 
we have explicitly derived this classical limit (proportional to 
$F_{\mu\nu}(0)$) in a general setting, and the complete agreement found 
in this work serves as a valuable cross-check for both types of calculations.  

With possible practical applications in mind, we have computed the 
proportionality constants $c^S$ and $c^T$ for a wide range of negative
mass parameter $\rho$. One notable feature of the $\rho$--dependence 
found is that $c^T(\rho)/\rho$ is surprisingly well described by a purely 
linear behavior in the region $0.2\le \rho \le 1.8$.

\vfill\eject

\begin{appendix}

\section{Alternative derivation to second order}
\label{app:secord}

The expansion of the overlap operator for constant fields to the second 
order in the lattice spacing can be simplified as follows. We start by 
noticing that
\begin{eqnarray}
   \nonumber
   \lim_{a\rightarrow 0} \frac{\partial^n}{\partial a^n} X &=& 
   \left( \sum_\mu A_\mu \frac{\partial}{\partial k_\mu} \right)^n X_0, \\
   \nonumber
   \lim_{a\rightarrow 0} \frac{\partial}{\partial a} B &=& 
   \sum_\mu A_\mu \frac{\partial}{\partial k_\mu} B_0, \\
   \lim_{a\rightarrow 0} \frac{\partial^2}{\partial a^2} B &=& 
   \left(\sum_\mu A_\mu \frac{\partial}{\partial k_\mu}\right)^2 B_0 + \delta_2,
\end{eqnarray}
where
\begin{equation}
   \delta_2 \,=\, \frac{1}{2} \sum_{\mu,\nu} \, [A_\mu,A_\nu] 
    \left( \frac{\partial X_0^\dagger}{\partial k_\mu}  
    \frac{\partial X_0}{\partial k_\nu}
   - \frac{\partial X_0^\dagger}{\partial k_\nu}  \frac{\partial X_0}
     {\partial k_\mu}\right)
\end{equation}
Using the above formulas and the straightforward algebra one can show that
\begin{eqnarray}
   \nonumber
   \lim_{a\rightarrow 0} \frac{\partial}{\partial a} \frac{1}{\rho} D &=& 
    \sum_\mu A_\mu \frac{\partial}{\partial k_\mu} X_0 B_0^{-1/2}, \\
    \lim_{a\rightarrow 0} \frac{\partial^2}{\partial a^2} \frac{1}{\rho} D &=& 
    \left(\sum_\mu A_\mu \frac{\partial}{\partial k_\mu} \right)^2
    X_0 B_0^{-1/2} -\, \frac{1}{2} X_0 B_0^{-3/2} \delta_2
\end{eqnarray}
The derivative terms $\,\sum_\mu A_\mu \frac{\partial}{\partial k_\mu} f(k)\,$ 
vanish upon momentum integration and we have
\begin{equation}
    \frac{1}{\rho} D_{0,0}\left(U(a)\right) \;=\; \frac{1}{\rho} 
    \,D_{0,0}(\identity) \,+\, \frac{a^2}{2}\,\frac{1}{(2\pi)^4} \, 
    \int d^4k \left(-\frac{1}{2} X_0 B_0^{-3/2} \delta_2\right) \,+\, \cO(a^3)
\end{equation}
After expanding $\delta_2$ and dropping the terms odd in $k_\mu$ we get
\begin{equation}
   \frac{1}{\rho} D_{0,0}\left(U(a)\right) \;=\; \frac{1}{\rho} 
   D_{0,0}(\identity) \,+\, 
   \frac{a^2}{2}\,  \frac{r}{(4r-\rho)^3} 
  \sum_{\mu \nu} \sigma_{\mu\nu} \times F_{\mu\nu} 
   \frac{1}{(2\pi)^4} \int d^4k  
            \frac{t_{\mu\nu}(k)}{B_0^{3/2}(k)}    
            \,+\, \cO(a^3)
\end{equation}
with $t_{\mu\nu}$ defined in Eq.~(\ref{eq:3.52}).

\vfill\eject

\section{Reduced Coefficients}
\label{app:coefs}

\def\boxit#1{\vbox{\hrule\hbox{\vrule\kern3pt
\vbox{\kern3pt#1\kern3pt}\kern3pt\vrule}\hrule}}

\def\qbox#1{
\setbox0 = \vtop{#1}
\ifdim \dp0 > 4pt \box0
\else \hbox{#1}
\fi
}

\long\def\coefftable[#1] #2
{
\setbox0=\vbox{
\halign{\strut##&\kern5pt\hfil##\hfil&\kern5pt\hfil##\hfil&
\kern5pt\hfil##\hfil&\kern5pt\hfil##\hfil&
\kern8pt{\qbox{\noindent##}}\hfil\cr
&\normalsize{$n_1$}&\normalsize{$n_2$}&\normalsize{$n_3$}&
\normalsize{$n_4$}&\normalsize{$c(n)/\rho$}\cr
\noalign{\vskip 2pt}\noalign{\hrule}\noalign{\vskip 3pt}
#2
\noalign{\vskip 3pt \hrule \vskip 3pt}
}}
\dimen0=\hsize
\advance\dimen0 by -\wd0
\divide\dimen0 by 2
\moveright \dimen0 \box0
\smallskip
\flushleft{#1}
\smallskip
}

The tables in this Appendix specify the reduced coefficients $c(n)$
and powers $j$ defining (via Eq.~(\ref{eq:3.76})) various constants 
appearing in the expansion of the scalar part of $D_{0,0}$ in 
classical lattice spacing. 
As discussed in Secs.~\ref{sec:scalar} and \ref{sec:finite_volume}, 
these coefficients can also be used for evaluation of corresponding 
constants on a finite symmetric torus with periodic boundary conditions 
in all directions.
\bigskip\bigskip

\footnotesize

\coefftable[\normalsize{Table 1 (part 1). Coefficients $c(n)$ for constants 
             $\alpha_{\mu\nu}$ of Eq.~(\ref{eq:3.80}). In this case $j=9/2$.}]
{
\input tex_trd_a4_j12_p_m4.5_aa_n
}

\vfill\eject

\coefftable[\normalsize{Table 1 (part 2). Coefficients $c(n)$ for constants 
            $\alpha_{\mu\nu}$ of Eq.~(\ref{eq:3.80}). In this case $j=9/2$.}]
{
\input tex_trd_a4_j12_p_m4.5_ab_n
}

\vfill\eject

\coefftable[\normalsize{Table 2 (part 1). Coefficients $c(n)$ for constants 
            $\beta_{\mu\nu}$ of Eq.~(\ref{eq:3.80}). In this case $j=9/2$.}]
{
\input tex_trd_a4_l12_p_m4.5_aa_n
}

\vfill\eject

\coefftable[\normalsize{Table 2 (part 2). Coefficients $c(n)$ for constants 
            $\beta_{\mu\nu}$ of Eq.~(\ref{eq:3.80}). In this case $j=9/2$.}]
{
\input tex_trd_a4_l12_p_m4.5_ab_n
}

\vfill\eject

\section{Sample Constant Field and Sample Points}
\label{app:constant}

\normalsize

The constant field $a_\mu$ used in numerical calculations of 
Sec.~\ref{sec:non-constant} (see Eq.~(\ref{eq:4.20})) is specified by
\medskip

$$ 
a_0 = \pmatrix{
0.333027 & 0.438189 + 0.318128 i& 0.374405 + 0.222978 i\cr
0.438189 - 0.318128 i& -0.154498 & 0.378804 + 0.22751 i\cr
0.374405 - 0.222978 i& 0.378804 - 0.22751 i& -0.178529 }
$$
\smallskip

$$
a_1 = \pmatrix{
0.230383  & 0.202411 +0.030092 i & 0.384110 -0.364239 i\cr
0.202411 -0.030092 i & -0.162558  & 0.448044 -0.138645 i\cr
0.384110 +0.364239 i & 0.448044 +0.138645 i & -0.067824 
}
$$
\smallskip

$$
a_2 = \pmatrix{
-0.254314  & 0.490871 -0.296415 i & 0.709724 -0.352729 i\cr
0.490871 +0.296415 i & 0.330503  & 0.970160 +0.109804 i\cr
0.709724 +0.352729 i & 0.970160 -0.109804 i & -0.076189 
}
$$
\smallskip

$$
a_3 = \pmatrix{
-0.055155  & 0.502826 +0.023544 i & 0.435721 -0.074859 i\cr
0.502826 -0.023544 i & 0.159857  & 0.581485 +0.032500 i\cr
0.435721 +0.074859 i & 0.581485 -0.032500 i & -0.104703 
}
$$
\medskip

The classical limit on non--constant backgrounds was evaluated 
at the following randomly generated points on the unit symmetric 
torus.
\medskip\medskip

{\tabskip =  .3cm
\halign to \hsize {
&\hfill#\hfill\cr
x & y &z & t \cr
0.103172983143503 & 0.9458719000957324 & 0.956116383180309 &
0.438505993255755 \cr
0.248194925172178 & 0.0893431271786154 & 0.336764036419274 &
0.118409829321844 \cr
0.174820791004537 & 0.7529966955146729 & 0.984186866641149 &
0.265965702089591 \cr
0.450532157114747 & 0.7958101168552333 & 0.274186255621803 &
0.321281009101127 \cr
0.416208702090845 & 0.6286435317521117 & 0.459165616384016 &
0.356014499573743 \cr
0.295893446617454 & 0.0590763301619042 & 0.678511201131179 &
0.781239724758651 \cr
0.192720463473952 & 0.1132044300661718 & 0.722394817950870 &
0.342733731502896 \cr
0.944525538301774 & 0.0238613028875564 & 0.385630781531596 &
0.224323902181052 \cr
0.769704747297237 & 0.2708646073728835 & 0.401443914890447 &
0.958358200091461 \cr
0.319172590182490 & 0.4750544905176503 & 0.127257659268644 &
0.637077190990334 \cr
}}

\end{appendix}

\vfill\eject

\end{document}
\bye

%% file: tex_trd_a4_j12_p_m4.5_aa_n
&0& 0& 0& 0& $48 \kappa^4 + 192 r^2 \kappa^4 + 1536 \kappa^6 - 1536 r^2 \kappa^6 - 10240 r^4 \kappa^6 + 
              12288 \kappa^8 - 73728 r^2 \kappa^8 + 122880 r^4 \kappa^8$\cr
&1& 0& 0& 0& $-32 r \kappa^3 - 1152 r \kappa^5 - 2560 r^3 \kappa^5 - 18432 r \kappa^7 - 
              24576 r^3 \kappa^7 + 245760 r^5 \kappa^7 - 131072 r \kappa^9 +$\cr
& & & & &    $589824 r^3 \kappa^9 - 983040 r^5 \kappa^9$\cr
&1& 1& 0& 0& $576 r^2 \kappa^4 + 12416 r^2 \kappa^6 - 30720 r^4 \kappa^6 - 14336 r^2 \kappa^8 + 
              516096 r^4 \kappa^8 - 1474560 r^6 \kappa^8$\cr
&1& 1& 1& 0& $-75264 r^3 \kappa^7 + 368640 r^5 \kappa^7 + 204800 r^3 \kappa^9 - 1179648 r^5 \kappa^9 + 
              1966080 r^7 \kappa^9$\cr
&1& 1& 1& 1& $-15360 r^4 \kappa^6 + 132096 r^4 \kappa^8 - 614400 r^6 \kappa^8$\cr
&2& 0& 0& 0& $-192 \kappa^4 - 96 r^2 \kappa^4 - 5760 \kappa^6 + 15232 r^2 \kappa^6 + 20480 r^4 \kappa^6 - 
              43008 \kappa^8 + 321536 r^2 \kappa^8 -$\cr
& & & & &    $319488 r^4 \kappa^8 - 491520 r^6 \kappa^8$\cr
&2& 1& 0& 0& $64 r \kappa^3 + 4032 r \kappa^5 + 1216 r^3 \kappa^5 + 74496 r \kappa^7 - 232704 r^3 \kappa^7 + 
              290816 r \kappa^9 - 1503232 r^3 \kappa^9 +$\cr
& & & & &    $1179648 r^5 \kappa^9 + 2949120 r^7 \kappa^9$\cr
&2& 1& 1& 0& $-1440 r^2 \kappa^4 - 13824 r^2 \kappa^6 - 42496 r^4 \kappa^6 - 198656 r^2 \kappa^8 + 
              952320 r^4 \kappa^8 - 1474560 r^6 \kappa^8$\cr
&2& 1& 1& 1& $6720 r^3 \kappa^5 - 18432 r^3 \kappa^7 + 242688 r^5 \kappa^7 + 77824 r^3 \kappa^9 - 
              774144 r^5 \kappa^9 + 1966080 r^7 \kappa^9$\cr
&2& 2& 0& 0& $240 \kappa^4 - 480 r^2 \kappa^4 + 8448 \kappa^6 - 25600 r^2 \kappa^6 - 23808 r^4 \kappa^6 + 
              77824 \kappa^8 - 561152 r^2 \kappa^8 + 861696 r^4 \kappa^8 +$\cr
& & & & &    $184320 r^6 \kappa^8$\cr
&2& 2& 1& 0& $-5568 r \kappa^5 + 16896 r^3 \kappa^5 - 78336 r \kappa^7 + 170496 r^3 \kappa^7 + 
              344064 r^5 \kappa^7 - 282624 r \kappa^9 + 1835008 r^3 \kappa^9 -$\cr
& & & & &    $3336192 r^5 \kappa^9 + 737280 r^7 \kappa^9$\cr
&2& 2& 1& 1& $24576 r^2 \kappa^6 - 92928 r^4 \kappa^6 + 110592 r^2 \kappa^8 - 12288 r^4 \kappa^8 - 
              1081344 r^6 \kappa^8$\cr
&2& 2& 2& 0& $-3840 \kappa^6 + 28416 r^2 \kappa^6 - 37888 r^4 \kappa^6 - 49152 \kappa^8 + 338944 r^2 \kappa^8 - 
              448512 r^4 \kappa^8 - 433152 r^6 \kappa^8$\cr
&2& 2& 2& 1& $23040 r \kappa^7 - 184320 r^3 \kappa^7 + 310272 r^5 \kappa^7 + 73728 r \kappa^9 - 
              518144 r^3 \kappa^9 + 516096 r^5 \kappa^9 + 1075200 r^7 \kappa^9$\cr
&2& 2& 2& 2& $7680 \kappa^8 - 67584 r^2 \kappa^8 + 241152 r^4 \kappa^8 - 224256 r^6 \kappa^8$\cr
&3& 0& 0& 0& $576 r \kappa^5 + 64 r^3 \kappa^5 + 11520 r \kappa^7 + 25344 r^3 \kappa^7 - 122880 r^5 \kappa^7 + 
              184320 r \kappa^9 - 765952 r^3 \kappa^9 +$\cr
& & & & &     $786432 r^5 \kappa^9 + 327680 r^7 \kappa^9$\cr
&3& 1& 0& 0& $-288 r^2 \kappa^4 - 20224 r^2 \kappa^6 + 27136 r^4 \kappa^6 + 38912 r^2 \kappa^8 - 
              509952 r^4 \kappa^8 + 983040 r^6 \kappa^8$\cr
&3& 1& 1& 0& $4800 r^3 \kappa^5 + 132096 r^3 \kappa^7 - 242688 r^5 \kappa^7 - 397312 r^3 \kappa^9 + 
              1306624 r^5 \kappa^9 - 983040 r^7 \kappa^9$\cr
&3& 1& 1& 1& $-15360 r^4 \kappa^6 - 153600 r^4 \kappa^8 + 307200 r^6 \kappa^8$\cr
&3& 2& 0& 0& $-2112 r \kappa^5 + 3776 r^3 \kappa^5 - 69888 r \kappa^7 + 116736 r^3 \kappa^7 + 
              76800 r^5 \kappa^7 - 540672 r \kappa^9 + 1982464 r^3 \kappa^9 -$\cr
& & & & &    $ 1628160 r^5 \kappa^9 - 1228800 r^7 \kappa^9$\cr
&3& 2& 1& 0& $44544 r^2 \kappa^6 - 105472 r^4 \kappa^6 + 409600 r^2 \kappa^8 - 657408 r^4 \kappa^8 - 188416 r^6 \kappa^8$\cr
&3& 2& 1& 1& $-156672 r^3 \kappa^7 + 420864 r^5 \kappa^7 - 260096 r^3 \kappa^9 + 233472 r^5 \kappa^9 + 131072 r^7 \kappa^9$\cr
&3& 2& 2& 0& $50688 r \kappa^7 - 238080 r^3 \kappa^7 + 258048 r^5 \kappa^7 + 432128 r \kappa^9 - 
              1789952 r^3 \kappa^9 + 1687552 r^5 \kappa^9 +$\cr
& & & & &    $792576 r^7 \kappa^9$\cr
&3& 2& 2& 1& $-211968 r^2 \kappa^8 + 1069056 r^4 \kappa^8 - 1422336 r^6 \kappa^8$\cr
&3& 2& 2& 2& $-73728 r \kappa^9 + 430080 r^3 \kappa^9 - 964608 r^5 \kappa^9 + 665600 r^7 \kappa^9$\cr
&3& 3& 0& 0& $8576 r^2 \kappa^6 - 11008 r^4 \kappa^6 - 18432 r^2 \kappa^8 + 132096 r^4 \kappa^8 - 184320 r^6 \kappa^8$\cr
&3& 3& 1& 0& $-130560 r^3 \kappa^7 + 205824 r^5 \kappa^7 + 188416 r^3 \kappa^9 - 634880 r^5 \kappa^9 + 401408 r^7 \kappa^9$\cr
&3& 3& 1& 1& $297984 r^4 \kappa^8 - 528384 r^6 \kappa^8$\cr
&3& 3& 2& 0& $-100352 r^2 \kappa^8 + 482304 r^4 \kappa^8 - 496640 r^6 \kappa^8$\cr
&3& 3& 2& 1& $237568 r^3 \kappa^9 - 1124352 r^5 \kappa^9 + 1382400 r^7 \kappa^9$\cr
&3& 3& 3& 0& $-20480 r^3 \kappa^9 - 75776 r^5 \kappa^9 + 133120 r^7 \kappa^9$\cr

%% file: tex_trd_a4_j12_p_m4.5_ab_n
&4& 0& 0& 0& $-384 \kappa^6 - 3328 r^2 \kappa^6 + 3328 r^4 \kappa^6 - 10752 \kappa^8 - 28672 r^2 \kappa^8 - 
              109056 r^4 \kappa^8 + 266240 r^6 \kappa^8$\cr
&4& 1& 0& 0& $384 r \kappa^5 - 192 r^3 \kappa^5 - 3072 r \kappa^7 + 140544 r^3 \kappa^7 - 168960 r^5 \kappa^7 + 
              111616 r \kappa^9 - 188416 r^3 \kappa^9 +$\cr
& & & & &    $885760 r^5 \kappa^9 - 1351680 r^7 \kappa^9$\cr
&4& 1& 1& 0& $-6912 r^2 \kappa^6 + 6912 r^4 \kappa^6 + 52224 r^2 \kappa^8 - 829440 r^4 \kappa^8 + 
              1019904 r^6 \kappa^8$\cr
&4& 1& 1& 1& $23040 r^3 \kappa^7 - 32256 r^5 \kappa^7 - 61440 r^3 \kappa^9 + 694272 r^5 \kappa^9 - 
              847872 r^7 \kappa^9$\cr
&4& 2& 0& 0& $2176 r^2 \kappa^6 - 3072 r^4 \kappa^6 - 9728 \kappa^8 + 251904 r^2 \kappa^8 - 430080 r^4 \kappa^8 + 
              64000 r^6 \kappa^8$\cr
&4& 2& 1& 0& $-4608 r \kappa^7 - 39936 r^3 \kappa^7 + 70656 r^5 \kappa^7 - 26624 r \kappa^9 - 770048 r^3 \kappa^9 + 
              2027520 r^5 \kappa^9 - 997376 r^7 \kappa^9$\cr
&4& 2& 1& 1& $23040 r^2 \kappa^8 + 79872 r^4 \kappa^8 - 155136 r^6 \kappa^8$\cr
&4& 2& 2& 0& $7680 \kappa^8 - 117760 r^2 \kappa^8 + 290304 r^4 \kappa^8 - 156672 r^6 \kappa^8$\cr
&4& 2& 2& 1& $3072 r \kappa^9 + 180224 r^3 \kappa^9 - 654336 r^5 \kappa^9 + 460800 r^7 \kappa^9$\cr
&4& 3& 0& 0& $19200 r \kappa^7 - 68352 r^3 \kappa^7 + 52224 r^5 \kappa^7 + 10240 r \kappa^9 - 33792 r^3 \kappa^9 - 
              182272 r^5 \kappa^9 + 332800 r^7 \kappa^9$\cr
&4& 3& 1& 0& $-144384 r^2 \kappa^8 + 651264 r^4 \kappa^8 - 592896 r^6 \kappa^8$\cr
&4& 3& 1& 1& $125952 r^3 \kappa^9 - 712704 r^5 \kappa^9 + 721920 r^7 \kappa^9$\cr
&4& 3& 2& 0& $-38912 r \kappa^9 + 417792 r^3 \kappa^9 - 950272 r^5 \kappa^9 + 563200 r^7 \kappa^9$\cr
&4& 4& 0& 0& $10240 \kappa^8 - 80384 r^2 \kappa^8 + 132096 r^4 \kappa^8 - 64512 r^6 \kappa^8$\cr
&4& 4& 1& 0& $-34816 r \kappa^9 + 303104 r^3 \kappa^9 - 595968 r^5 \kappa^9 + 337920 r^7 \kappa^9$\cr
&5& 0& 0& 0& $6912 r \kappa^7 - 768 r^3 \kappa^7 - 6144 r^5 \kappa^7 - 21504 r \kappa^9 + 131072 r^3 \kappa^9 + 
              7168 r^5 \kappa^9 - 188416 r^7 \kappa^9$\cr
&5& 1& 0& 0& $-3072 r^2 \kappa^6 + 2816 r^4 \kappa^6 - 25600 r^2 \kappa^8 - 181248 r^4 \kappa^8 + 
              237568 r^6 \kappa^8$\cr
&5& 1& 1& 0& $41472 r^3 \kappa^7 - 41472 r^5 \kappa^7 - 63488 r^3 \kappa^9 + 706560 r^5 \kappa^9 - 
              765952 r^7 \kappa^9$\cr
&5& 1& 1& 1& $-92160 r^4 \kappa^8 + 98304 r^6 \kappa^8$\cr
&5& 2& 0& 0& $-9216 r \kappa^7 + 21504 r^3 \kappa^7 - 10752 r^5 \kappa^7 + 115712 r \kappa^9 - 
              458752 r^3 \kappa^9 + 489472 r^5 \kappa^9 - 64512 r^7 \kappa^9$\cr
&5& 2& 1& 0& $64512 r^2 \kappa^8 - 196608 r^4 \kappa^8 + 107520 r^6 \kappa^8$\cr
&5& 2& 1& 1& $-64512 r^3 \kappa^9 + 251904 r^5 \kappa^9 - 162816 r^7 \kappa^9$\cr
&5& 2& 2& 0& $-52224 r \kappa^9 + 137216 r^3 \kappa^9 - 58368 r^5 \kappa^9 - 49152 r^7 \kappa^9$\cr
&5& 3& 0& 0& $-6656 r^2 \kappa^8 + 46080 r^4 \kappa^8 - 42496 r^6 \kappa^8$\cr
&5& 3& 1& 0& $63488 r^3 \kappa^9 - 258048 r^5 \kappa^9 + 247808 r^7 \kappa^9$\cr
&5& 4& 0& 0& $-41984 r \kappa^9 + 156672 r^3 \kappa^9 - 189440 r^5 \kappa^9 + 76800 r^7 \kappa^9$\cr
&6& 0& 0& 0& $4608 \kappa^8 - 25600 r^2 \kappa^8 + 24576 r^4 \kappa^8 - 3584 r^6 \kappa^8$\cr
&6& 1& 0& 0& $-2304 r \kappa^7 + 6912 r^3 \kappa^7 - 4608 r^5 \kappa^7 - 35840 r \kappa^9 + 112640 r^3 \kappa^9 - 
              10240 r^5 \kappa^9 - 66560 r^7 \kappa^9$\cr
&6& 1& 1& 0& $19968 r^2 \kappa^8 - 64512 r^4 \kappa^8 + 44544 r^6 \kappa^8$\cr
&6& 1& 1& 1& $-21504 r^3 \kappa^9 + 73728 r^5 \kappa^9 - 52224 r^7 \kappa^9$\cr
&6& 2& 0& 0& $-7680 \kappa^8 + 42496 r^2 \kappa^8 - 58368 r^4 \kappa^8 + 23552 r^6 \kappa^8$\cr
&6& 2& 1& 0& $34816 r \kappa^9 - 182272 r^3 \kappa^9 + 258048 r^5 \kappa^9 - 110592 r^7 \kappa^9$\cr
&6& 3& 0& 0& $23552 r \kappa^9 - 62464 r^3 \kappa^9 + 37888 r^5 \kappa^9 + 1024 r^7 \kappa^9$\cr
&7& 0& 0& 0& $-4096 r \kappa^9 + 17408 r^3 \kappa^9 - 22528 r^5 \kappa^9 + 9216 r^7 \kappa^9$\cr
&7& 1& 0& 0& $1536 r^2 \kappa^8 - 3072 r^4 \kappa^8 + 1536 r^6 \kappa^8$\cr
&7& 1& 1& 0& $-9216 r^3 \kappa^9 + 18432 r^5 \kappa^9 - 9216 r^7 \kappa^9$\cr
&7& 2& 0& 0& $5120 r \kappa^9 - 23552 r^3 \kappa^9 + 31744 r^5 \kappa^9 - 13312 r^7 \kappa^9$\cr
&8& 1& 0& 0& $-1024 r \kappa^9 + 3072 r^3 \kappa^9 - 3072 r^5 \kappa^9 + 1024 r^7 \kappa^9$\cr

%% file: tex_trd_a4_l12_p_m4.5_aa_n
&0& 0& 0& 0& $-5120 r^4 \kappa^6 + 61440 r^4 \kappa^8$\cr
&1& 0& 0& 0& $1024 r^3 \kappa^5 - 49152 r^3 \kappa^7 + 122880 r^5 \kappa^7 + 98304 r^3 \kappa^9 - 491520 r^5 \kappa^9$\cr
&1& 1& 0& 0& $4480 r^2 \kappa^6 - 24576 r^4 \kappa^6 - 71680 r^2 \kappa^8 + 589824 r^4 \kappa^8 - 737280 r^6 \kappa^8$\cr
&1& 1& 1& 0& $-53760 r^3 \kappa^7 + 147456 r^5 \kappa^7 + 286720 r^3 \kappa^9 - 1179648 r^5 \kappa^9 + 983040 r^7 \kappa^9$\cr
&1& 1& 1& 1& $107520 r^4 \kappa^8 - 196608 r^6 \kappa^8$\cr
&2& 0& 0& 0& $-96 r^2 \kappa^4 + 128 r^2 \kappa^6 + 7168 r^4 \kappa^6 - 45056 r^2 \kappa^8 - 49152 r^4 \kappa^8 - 
              245760 r^6 \kappa^8$\cr
&2& 1& 0& 0& $1856 r^3 \kappa^5 + 15360 r \kappa^7 + 29952 r^3 \kappa^7 - 55296 r^5 \kappa^7 - 40960 r \kappa^9 + 
              311296 r^3 \kappa^9 - 294912 r^5 \kappa^9 +$\cr
& & & & &    $1474560 r^7 \kappa^9$\cr
&2& 1& 1& 0& $-6656 r^4 \kappa^6 - 87040 r^2 \kappa^8 - 141312 r^4 \kappa^8 - 73728 r^6 \kappa^8$\cr
&2& 1& 1& 1& $-3072 r^5 \kappa^7 + 51200 r^3 \kappa^9 + 36864 r^5 \kappa^9 + 245760 r^7 \kappa^9$\cr
&2& 2& 0& 0& $-48 \kappa^4 + 96 r^2 \kappa^4 - 1536 \kappa^6 + 2176 r^2 \kappa^6 - 16128 r^4 \kappa^6 - 3328 \kappa^8 + 
              35840 r^2 \kappa^8 + 59904 r^4 \kappa^8 +$\cr
& & & & &    $258048 r^6 \kappa^8$\cr
&2& 2& 1& 0& $960 r \kappa^5 - 1920 r^3 \kappa^5 + 18432 r \kappa^7 - 50688 r^3 \kappa^7 + 144384 r^5 \kappa^7 + 
              33792 r \kappa^9 - 28672 r^3 \kappa^9 +$\cr
& & & & &    $18432 r^5 \kappa^9 - 737280 r^7 \kappa^9$\cr
&2& 2& 1& 1& $-3840 r^2 \kappa^6 + 7680 r^4 \kappa^6 - 36864 r^2 \kappa^8 + 150528 r^4 \kappa^8 - 190464 r^6 \kappa^8$\cr
&2& 2& 2& 0& $768 \kappa^6 - 5376 r^2 \kappa^6 + 10240 r^4 \kappa^6 + 12288 \kappa^8 - 80896 r^2 \kappa^8 + 43008 r^4 \kappa^8 - 
              248832 r^6 \kappa^8$\cr
&2& 2& 2& 1& $-4608 r \kappa^7 + 32256 r^3 \kappa^7 - 61440 r^5 \kappa^7 - 24576 r \kappa^9 + 149504 r^3 \kappa^9 - 
              184320 r^5 \kappa^9 + 399360 r^7 \kappa^9$\cr
&2& 2& 2& 2& $-1536 \kappa^8 + 12288 r^2 \kappa^8 - 33792 r^4 \kappa^8 + 76800 r^6 \kappa^8$\cr
&3& 0& 0& 0& $-64 r^3 \kappa^5 + 60672 r^3 \kappa^7 - 67584 r^5 \kappa^7 - 57344 r^3 \kappa^9 + 294912 r^5 \kappa^9 + 
              163840 r^7 \kappa^9$\cr
&3& 1& 0& 0& $-8960 r^2 \kappa^6 + 9728 r^4 \kappa^6 + 117760 r^2 \kappa^8 - 964608 r^4 \kappa^8 + 712704 r^6 \kappa^8$\cr
&3& 1& 1& 0& $107520 r^3 \kappa^7 - 107520 r^5 \kappa^7 - 542720 r^3 \kappa^9 + 2330624 r^5 \kappa^9 - 1228800 r^7 \kappa^9$\cr
&3& 1& 1& 1& $-215040 r^4 \kappa^8 + 208896 r^6 \kappa^8$\cr
&3& 2& 0& 0& $960 r \kappa^5 - 896 r^3 \kappa^5 + 3072 r \kappa^7 - 70656 r^3 \kappa^7 + 98304 r^5 \kappa^7 + 
              64512 r \kappa^9 - 231424 r^3 \kappa^9 +$\cr
& & & & &    $374784 r^5 \kappa^9 - 983040 r^7 \kappa^9$\cr
&3& 2& 1& 0& $-15360 r^2 \kappa^6 + 14336 r^4 \kappa^6 - 96256 r^2 \kappa^8 + 546816 r^4 \kappa^8 - 524288 r^6 \kappa^8$\cr
&3& 2& 1& 1& $46080 r^3 \kappa^7 - 43008 r^5 \kappa^7 + 167936 r^3 \kappa^9 - 528384 r^5 \kappa^9 + 262144 r^7 \kappa^9$\cr
&3& 2& 2& 0& $-13824 r \kappa^7 + 105984 r^3 \kappa^7 - 104448 r^5 \kappa^7 - 94208 r \kappa^9 + 296960 r^3 \kappa^9 - 
              446464 r^5 \kappa^9 +$\cr
& & & & &    $878592 r^7 \kappa^9$\cr
&3& 2& 2& 1& $55296 r^2 \kappa^8 - 423936 r^4 \kappa^8 + 417792 r^6 \kappa^8$\cr
&3& 2& 2& 2& $12288 r \kappa^9 - 73728 r^3 \kappa^9 + 282624 r^5 \kappa^9 - 352256 r^7 \kappa^9$\cr
&3& 3& 0& 0& $640 r^2 \kappa^6 - 512 r^4 \kappa^6 - 82944 r^2 \kappa^8 + 337920 r^4 \kappa^8 - 202752 r^6 \kappa^8$\cr
&3& 3& 1& 0& $-7680 r^3 \kappa^7 + 6144 r^5 \kappa^7 + 475136 r^3 \kappa^9 - 1404928 r^5 \kappa^9 + 679936 r^7 \kappa^9$\cr
&3& 3& 1& 1& $15360 r^4 \kappa^8 - 12288 r^6 \kappa^8$\cr
&3& 3& 2& 0& $91136 r^2 \kappa^8 - 353280 r^4 \kappa^8 + 229376 r^6 \kappa^8$\cr
&3& 3& 2& 1& $-182272 r^3 \kappa^9 + 706560 r^5 \kappa^9 - 458752 r^7 \kappa^9$\cr
&3& 3& 3& 0& $-108544 r^3 \kappa^9 + 260096 r^5 \kappa^9 - 114688 r^7 \kappa^9$\cr

%% file: tex_trd_a4_l12_p_m4.5_ab_n
&4& 0& 0& 0& $-2432 r^2 \kappa^6 + 1792 r^4 \kappa^6 + 37888 r^2 \kappa^8 - 167424 r^4 \kappa^8 + 151552 r^6 \kappa^8$\cr
&4& 1& 0& 0& $-15360 r \kappa^7 + 85248 r^3 \kappa^7 - 56832 r^5 \kappa^7 + 51200 r \kappa^9 - 456704 r^3 \kappa^9 + 
              1276928 r^5 \kappa^9 -$\cr
& & & & &    $860160 r^7 \kappa^9$\cr
&4& 1& 1& 0& $122880 r^2 \kappa^8 - 506880 r^4 \kappa^8 + 325632 r^6 \kappa^8$\cr
&4& 1& 1& 1& $-122880 r^3 \kappa^9 + 448512 r^5 \kappa^9 - 282624 r^7 \kappa^9$\cr
&4& 2& 0& 0& $768 \kappa^6 - 2176 r^2 \kappa^6 + 2048 r^4 \kappa^6 - 5632 \kappa^8 + 1536 r^2 \kappa^8 + 61440 r^4 \kappa^8 - 
              134656 r^6 \kappa^8$\cr
&4& 2& 1& 0& $-9216 r \kappa^7 + 26112 r^3 \kappa^7 - 24576 r^5 \kappa^7 + 2048 r \kappa^9 - 22528 r^3 \kappa^9 - 
              110592 r^5 \kappa^9 + 309248 r^7 \kappa^9$\cr
&4& 2& 1& 1& $18432 r^2 \kappa^8 - 52224 r^4 \kappa^8 + 49152 r^6 \kappa^8$\cr
&4& 2& 2& 0& $-7680 \kappa^8 + 71680 r^2 \kappa^8 - 175104 r^4 \kappa^8 + 156672 r^6 \kappa^8$\cr
&4& 2& 2& 1& $15360 r \kappa^9 - 143360 r^3 \kappa^9 + 350208 r^5 \kappa^9 - 313344 r^7 \kappa^9$\cr
&4& 3& 0& 0& $10752 r \kappa^7 - 28416 r^3 \kappa^7 + 12288 r^5 \kappa^7 - 50176 r \kappa^9 + 347136 r^3 \kappa^9 - 
              727040 r^5 \kappa^9 + 404480 r^7 \kappa^9$\cr
&4& 3& 1& 0& $-86016 r^2 \kappa^8 + 227328 r^4 \kappa^8 - 98304 r^6 \kappa^8$\cr
&4& 3& 1& 1& $86016 r^3 \kappa^9 - 227328 r^5 \kappa^9 + 98304 r^7 \kappa^9$\cr
&4& 3& 2& 0& $51200 r \kappa^9 - 245760 r^3 \kappa^9 + 495616 r^5 \kappa^9 - 331776 r^7 \kappa^9$\cr
&4& 4& 0& 0& $7424 \kappa^8 - 44032 r^2 \kappa^8 + 62976 r^4 \kappa^8 - 21504 r^6 \kappa^8$\cr
&4& 4& 1& 0& $-29696 r \kappa^9 + 176128 r^3 \kappa^9 - 251904 r^5 \kappa^9 + 86016 r^7 \kappa^9$\cr
&5& 0& 0& 0& $2304 r^3 \kappa^7 - 1536 r^5 \kappa^7 - 20480 r^3 \kappa^9 + 91136 r^5 \kappa^9 - 106496 r^7 \kappa^9$\cr
&5& 1& 0& 0& $25600 r^2 \kappa^8 - 86016 r^4 \kappa^8 + 51200 r^6 \kappa^8$\cr
&5& 1& 1& 0& $-102400 r^3 \kappa^9 + 288768 r^5 \kappa^9 - 167936 r^7 \kappa^9$\cr
&5& 2& 0& 0& $-4608 r \kappa^7 + 9984 r^3 \kappa^7 - 6144 r^5 \kappa^7 + 1024 r \kappa^9 - 11264 r^3 \kappa^9 - 
              22528 r^5 \kappa^9 + 113664 r^7 \kappa^9$\cr
&5& 2& 1& 0& $36864 r^2 \kappa^8 - 79872 r^4 \kappa^8 + 49152 r^6 \kappa^8$\cr
&5& 2& 1& 1& $-36864 r^3 \kappa^9 + 79872 r^5 \kappa^9 - 49152 r^7 \kappa^9$\cr
&5& 2& 2& 0& $15360 r \kappa^9 - 81920 r^3 \kappa^9 + 141312 r^5 \kappa^9 - 116736 r^7 \kappa^9$\cr
&5& 3& 0& 0& $-7168 r^2 \kappa^8 + 18432 r^4 \kappa^8 - 8192 r^6 \kappa^8$\cr
&5& 3& 1& 0& $28672 r^3 \kappa^9 - 73728 r^5 \kappa^9 + 32768 r^7 \kappa^9$\cr
&5& 4& 0& 0& $-19456 r \kappa^9 + 67584 r^3 \kappa^9 - 71680 r^5 \kappa^9 + 20480 r^7 \kappa^9$\cr
&6& 0& 0& 0& $-6656 r^2 \kappa^8 + 12288 r^4 \kappa^8 - 5632 r^6 \kappa^8$\cr
&6& 1& 0& 0& $-10240 r \kappa^9 + 44032 r^3 \kappa^9 - 51200 r^5 \kappa^9 + 17408 r^7 \kappa^9$\cr
&6& 2& 0& 0& $-1536 \kappa^8 + 17408 r^2 \kappa^8 - 29184 r^4 \kappa^8 + 13312 r^6 \kappa^8$\cr
&6& 2& 1& 0& $6144 r \kappa^9 - 69632 r^3 \kappa^9 + 116736 r^5 \kappa^9 - 53248 r^7 \kappa^9$\cr
&6& 3& 0& 0& $13312 r \kappa^9 - 38912 r^3 \kappa^9 + 35840 r^5 \kappa^9 - 10240 r^7 \kappa^9$\cr
&7& 0& 0& 0& $7168 r^3 \kappa^9 - 14336 r^5 \kappa^9 + 7168 r^7 \kappa^9$\cr
&7& 2& 0& 0& $3072 r \kappa^9 - 16384 r^3 \kappa^9 + 23552 r^5 \kappa^9 - 10240 r^7 \kappa^9$\cr